\documentclass[a4paper, twoside, 11pt]{report}

\usepackage[T1]{fontenc}
\usepackage{lmodern}
\usepackage[sc]{mathpazo}
\usepackage{graphicx}
\usepackage{amsmath}
\usepackage{pdfsync}

\usepackage{booktabs}

\usepackage[labelfont={sf,bf}, textfont=sf]{subfig}

\usepackage{caption}
\usepackage{tabularx}

\begin{document}
\title{	\vspace{-1cm}\textrm{\huge{\textsc{Cross-Frequency Coupling of Neuronal Oscillations During Cognition}}}\\
		\vspace{6mm} 
		\sc{\Large{Thomas E. Gorochowski}\\
		\vspace{14.5cm}}}

\author{
	\large{Supervisors: Dr. Rafal Bogacz, Dr. Matthew Jones}
	\vspace{0.15in} \\
	\large{\textsc{Bristol Centre for Complexity Sciences}} \\ 
	\large{\textsc{University of Bristol}} \\ 
	\small{2009}}
\date{}

\maketitle

\pagestyle{empty}
~
\pagenumbering{roman}
\clearpage
\pagestyle{plain}

\begin{abstract}
How the brain co-ordinates the actions of distant regions in an efficient manner is an open problem. Many believe that cross-frequency coupling between the amplitude of high frequency local field potential oscillations in one region and the phase of lower frequency signals in another may form a possible mechanism. This work provides a preliminary study from both an experimental and theoretical viewpoint, concentrating on possible coupling between the hippocampus and pre-frontal cortex in rats during tasks involving working memory, spatial navigation and decision making processes. Attempts to search for such coupling events are made using newly developed MATLAB scripts. This leads to the discovery of increased envelope-to-signal correlation (ESC) between the 1-10Hz hippocampus theta and 30-40Hz pre-fontal cortex gamma bands when a choice turn is approached during a T-maze task. From a theoretical perspective, a standard connectionist modelling approach is extended to allow for the formation of oscillations. Although detrimental to overall average task performance, this did lead to a reduced increase in performance variation as noise was increased, when compared to a standard non-oscillating model.
\end{abstract}

\cleardoublepage
\tableofcontents
\cleardoublepage

\pagenumbering{arabic}
\setcounter{page}{1}

\cleardoublepage

\chapter{Introduction}

\section{Brain Structure and Oscillations}

The brain comprises of approximately $10^{11}$ neurones that are connected through around $10^{14}$ synapses. The topology of the network generated by these connections is non-random, with the brain containing distinct regions that can be categorised according to the function they fulfil. Understanding exactly what these functions are is an open problem, however, specific regions have been extensively studied with some success. Two such regions are the Hippocampus (HPC) and Pre-frontal Cortex (PFC). The hippocampus has been shown to be related to memory encoding, spatial memory and navigation, while the pre-frontal cortex 
is thought to be involved in working memory, planning of complex behaviours, decision making and co-ordination of thought in relation to internal goals.

Having a brain consisting of functionally distinct areas raises many questions when considering how these work together in parallel for tasks requiring a multitude of different functions. With the transmission of information from one neurone to another taking a measurable period of time, it is necessary that distant regions become co-ordinated to ensure that they process the same information together. The idea of many brain regions working towards some conscious goal is what some term a ``cognitive moment''. The method of co-ordinating this behaviour is still unknown, however, it is believed that oscillations may play a key role \cite{Singer:1999p7559,Varela:2001p7792,Buzsaki:2006}.

Oscillations in the brain arise naturally due to the interplay between excitatory and inhibitory neurones. The connectivity of these different neurone types in feedback, feed-forward and lateral inhibition structures permits for non-linear relationships to be formed, causing initial instabilities to `seed' oscillatory behaviour in groups of neurones. In addition to structural causes, they are also believed to be induced as a mechanism by which distant regions can become co-ordinated for the purposes of a task at hand. Common frequencies that are observed at various locations within the brain are shown in Table \ref{tab:FreqBands}.

\begin{table}[tbp]
\centering
\begin{tabular}{ l c l}
	\toprule
	\textbf{Band Name} & \textbf{Frequency} (Hz) & \textbf{Description} \\ 
	\midrule 
	Theta ($\theta$) & $4 - 10$ &  Learning and memory retrieval\\
	Alpha/Mu ($\alpha$) & $8 - 15$ &  Visual cortex in idle state (eyes closed)\\
	Beta ($\beta$) & $15 - 30$ &  Active thinking \\
	Gamma ($\gamma$) & $30 - 90$ & Local structure co-ordination \\
	\bottomrule
\end{tabular}
\caption{\textbf{Standard Frequency Bands}}\label{tab:FreqBands}
\end{table}

One of the possible ways in which these oscillations could assist co-ordination is via cross-frequency coupling. This is where properties of two different frequency signals become correlated. For the purpose of this project we will explicitly be looking at phase to amplitude coupling. This is where the amplitude of a high frequency signal is modulated by the phase of a lower frequency. If the source of the low and high frequencies is different, then changes in amplitude of the high frequency carrier can be used as a mechanism of transferring timing information between the two regions. 

Having co-ordinated behaviour would only be necessary when the function of both regions is required for the task at hand. Recent work supports this idea \cite{Tort:2008}, showing a task dependance on the formation of similar coupling between the striatum and hippocampus via the theta and gamma frequency bands. Also, work presented in \cite{Jones:2005p7436} shows task dependant theta frequency coupling between the hippocampus and pre-frontal cortex. Due to these findings we will concentrate on seeking relationships between theta and gamma frequency bands.

\section{Experimental Configuration}

\begin{figure}[tbp]
\centering
\includegraphics[width=5.5in]{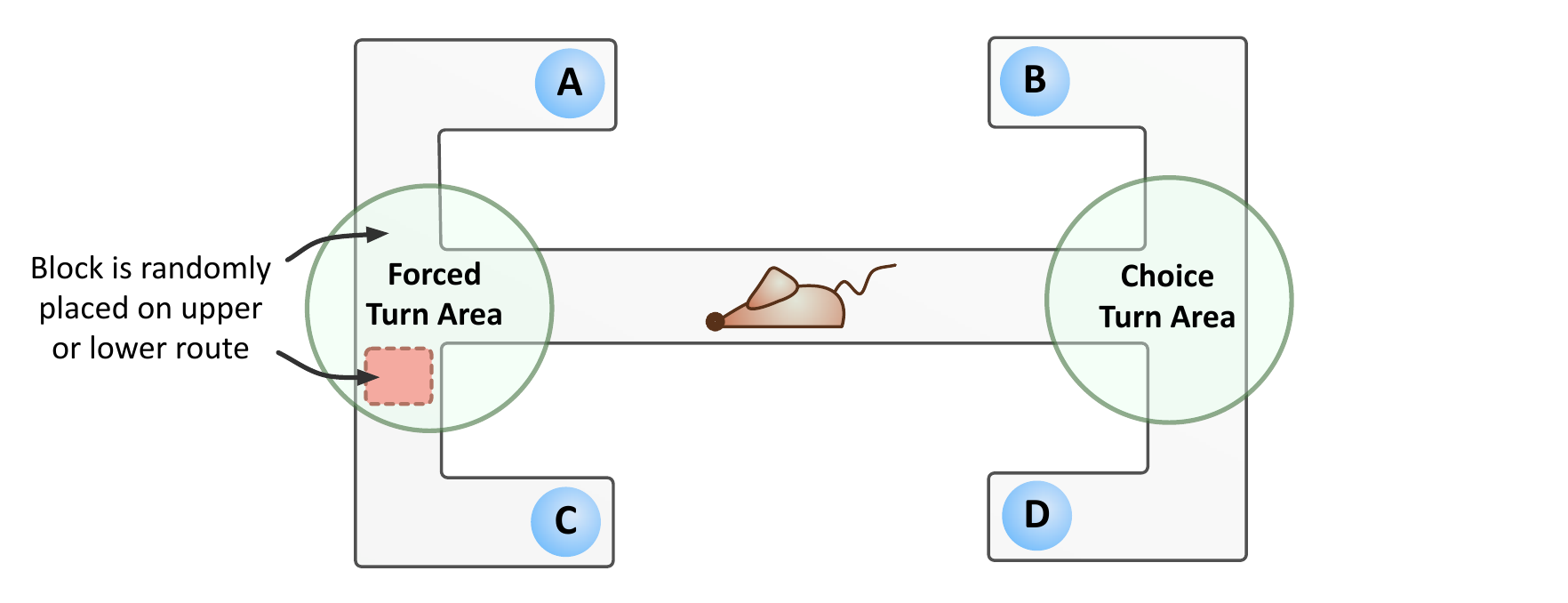}
\caption{\textbf{T-Maze Design} - Each arm of the maze has been labelled and areas of interest highlighted.}\label{fig:T-Maze}
\end{figure}

To investigate if cross-frequency coupling occurs during ``cognitive moments'' a T-maze task was selected, using rats as the experimental subjects. The maze design is shown in Figure \ref{fig:T-Maze} with the maze containing 4 arms, labelled A, B, C, D, and a block forcing the rat to take a route via arm A or C. For the rat to complete a trial successfully, and receive a reward, it must remember the previous forced direction and use that information to inform its decision when presented with a T-junction at the choice end of the maze. For example, if the rat has previously been forced to take the route towards arm A, then on reaching the next choice turn the correct option is the route towards arm B. In contrast, if forced towards arm C then the correct choice is a right hand turn towards arm D. Rats can be trained on such a task to reach a success rate of around 90\% with a standard deviation of 8\%.

This type of task is suitable for analysing cross-frequency coupling because for a high accuracy to be achieved it is necessary for the rat to integrate several different types of information. Firstly, a form of working memory is required to hold the previous forced turn direction. Next, the rat will likely need an understanding of it's spatial environment to allow for a choice turn to be anticipated, providing additional time in which to make the decision. Finally, both these pieces of information must be analysed together to make a decision on the course of action to take. For these reasons, we expected to see a greater coupling between the HPC and PFC as the rat approaches the choice turn. This is in contrast to the forced turn which does not require the integration of such information and allows it to be used as a control.

\section{Project Description}
This project attempts to analyse actual brain recordings to search of possible cross-frequency coupling events and investigates ways in which oscillations can be incorporated into existing modelling approaches. These aims fall into three main areas:
\begin{enumerate}
	\item \textbf{Detection of Cross-Frequency Coupling} -
	The first part of the project analyses data recorded from implanted electrodes in rats that perform a T-maze task. Methods to extract the amplitude of a signal are developed and an analysis of the correlation between different frequency bands is performed. Attempts are made to reduce user involvement through automation where possible.
	
	\item \textbf{Connectionist Model of Interactions} - 
	A standard way of modelling neuronal networks is to use a connectionist approach. We develop a model using this methodology with the aim of understanding which aspects  are important and if certain structures and connections are necessary for efficient execution of a given task, e.g. a decision based on historical information. At this stage no oscillatory features are considered, instead the focus is on producing a baseline model that can be used for comparison later.
	
	\item \textbf{Incorporating Oscillations} - 
	The final part of the project considers the effect of oscillations. Using the basic connectionist model developed previously as a starting point, we attempt to develop a method of incorporating oscillations and assess possible reasons they may be beneficial.
\end{enumerate}

\section{Report Structure and Contents}
This report is structured into 3 main parts. First, chapter 2 outlines the methods used to analyse actual brain recordings and the models developed to understand the basis for the formation of oscillations. Next, chapter 3 presents results from the analysis and modelling, with a discussion on the importance of any findings in relation to the initial project goals. Finally, chapter 4 summarises the results, evaluating the success of the project and provides possible future directions in which the work could be taken.

Due to space constraints, code listings have not been included in this report. Instead, all code is available on request by e-mailing \texttt{thomas.gorochowski@bristol.ac.uk}.

\cleardoublepage
\chapter{Methods}

In order to gain an understanding of any cross-frequency coupling present and a possible basis for its formation, the project was approached in two different ways. The first considered actual brain activity with the aim of understanding where cross-frequency coupling occurs, the key frequency bands and any behavioural aspects that lead to its formation. The second concentrated on gaining intuition as to why this coupling may occur through the incorporation of oscillations into existing connectionist modelling approaches. The following chapter explains how data was collected and describes the methods used during analysis and modelling.

\section{Data Collection}
Data from experiments was collected using a Neuralynx recording system. This allows for simultaneous recordings from electrodes at user specified sampling rates. All data analysed in this report used a sampling rate of 4069Hz and where necessary was down-sampled to 500Hz to help reduce the processing time of analyses. During the recording process two file types are generated:
\begin{itemize}
	\item \textbf{Continuous Recording Data (NCS)} - These are direct recordings of the Local Field Potential (LFP) for each of the implanted electrodes. The LFP measures the average activity of input signals into a region, being thought to be specifically related to the changes at dendritic synapses. Many consider this as the synchronised input into the region of interest.  
	
	\item \textbf{Spike Data (NTT)} - This data is also recorded from the same electrodes but using higher frequency components and extracted from timing differences in the wires that make up the recording device. To single out individual neurone spike trains, post-processing of the data is required to cluster related events. Spike data is thought to relate to the output activity of a region.  
\end{itemize}

In addition to these recordings, videos were taken of the rat during trails and rest periods to permit the inclusion of behavioural aspects into any analyses. Software exists to automatically extract the location of the rat in the T-maze during a trial from this data. Having this information makes it possible to assess if different areas in the maze (e.g. choice junction) are linked to differences in brain activity.

It should be made clear that all data used in this report was provided by Matthew Jones' lab. No physical recordings were performed as part of this work.

\section{Data Analysis}

\begin{figure}[tbp]
\centering
\includegraphics[width=6.3in]{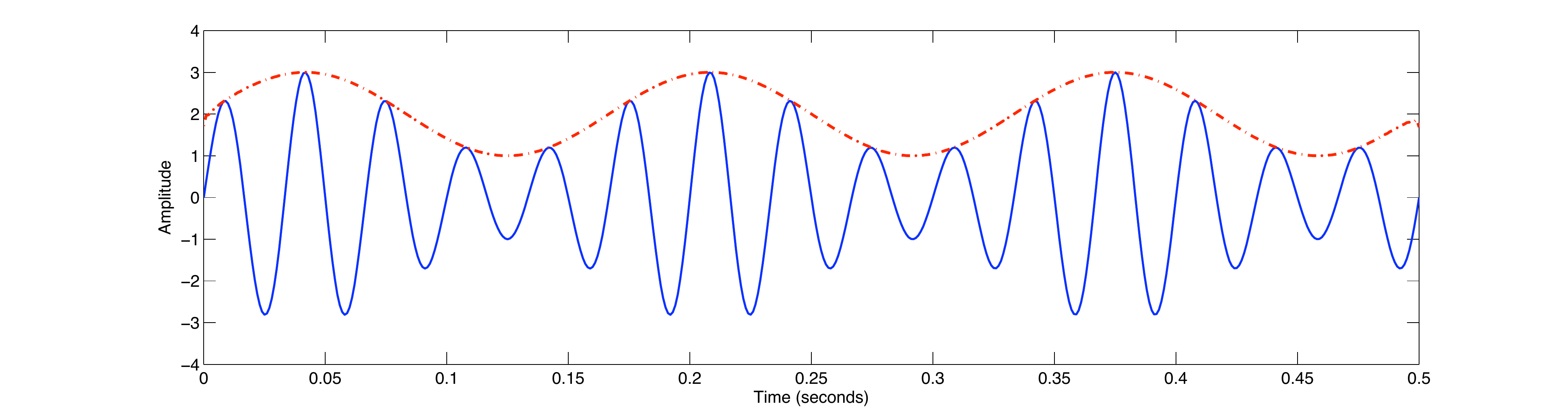}
\caption{\textbf{Phase to Amplitude Cross Frequency Coupling} - This illustrates how the power of a 30Hz carrier signal (blue solid line) can be modulated by a slower 6Hz signal (red dotted line). In the brain it is thought the region of each signal would be different with coupling used as a possible way to transfer information or co-ordinate activity.}\label{fig:PowPhaCouple}
\end{figure}

When considering cross-frequency coupling between signals there are many possible forms. These include phase-amplitude, amplitude-amplitude, frequency-amplitude and phase-phase. This report will focus on the discovery of phase-amplitude coupling, as shown in Figure \ref{fig:PowPhaCouple}.

\subsection{Loading Data}
All analysis was carried out on continuous recordings (NCS data) using MATLAB on Mac OS X. Experimental data was recorded by a proprietary system from Neuralynx that generates binary files containing the actual recordings. Tools are provided to both load and down sample these files for use with MATLAB, however, these unfortunately are only available for the Windows platform. As an alternative, the FieldTrip toolkit (\texttt{http://www.ru.nl/fcdonders/fieldtrip/}) was found, providing pure MATLAB implementations to load Neuralynx and several other formats directly.

The FieldTrip routines extract `raw' data from the binary files meaning further post-processing is required for a suitable structure to be generated. The raw data is in the form of a matrix of size $n \times 512$, where $n$ is the number of buffers recorded and $512$ is the size the buffer (\texttt{uint64} value for each reading). Due to this structure, time stamp data is only held for the first entries in the buffer. To convert this data, the matrix is unravelled and time stamps interpolated for intermediate readings. This is carried out automatically by the \texttt{process\_ncs.m} script developed to carry out similar conversions to those used by Neuralynx. This script produces 3 outputs, a vector of recorded LFP values, a vector of time stamps and the sampling frequency of the data.

\subsection{Power Spectrum}
The power spectrum breaks up a signal into its constituent frequency components. These are normally computed by performing a Fast Fourier Transform (FFT) on the signal and using the generated coefficients. These contain the the absolute values of the information content at each frequency and generally the square of these is taken for use in the power spectrum.

Power spectrum were used extensively in this project to help detect likely areas in which frequency-coupling may be occurring. It has been shown previously \cite{Tort:2008} that task dependant coupling is seen between the striatum and hippocampus, with theta modulating gamma amplitude. As a first step to finding similar task dependant relationships a method was developed to inspect how the power spectrum of specific frequency bands altered over time. A threshold was selected for each frequency band and areas where the power of both bands exceeded the threshold were recorded. By varying the frequency bands and thresholds, specific time periods of interest could be highlighted. It was hoped these would then relate to specific areas of the T-maze, e.g. decision point, and could be selected as a starting point for more detailed analyses.

To carry out this process two scripts were developed, the first \texttt{pow\_spec\_band.m}, extracts the power of a signal in a given frequency band. The second \texttt{comp\_2sig\_band\_pow.m}, takes two input signals, frequency bands and thresholds, and performs a comparison of the two outputting a boolean value at each time point where both frequency bands have a power in excess of the specified thresholds. 

Both scripts make use of the Chronux MATLAB library which had been developed specifically for use by neuroscientists to assist in frequency domain analysis and data fitting tasks. Further information regarding the library can be found at \texttt{http://chronux.org}.

\subsection{Phase and Amplitude Decomposition}\label{sec:ampDec}

\begin{figure}[tbp]
\centering
\includegraphics[width=4.4in]{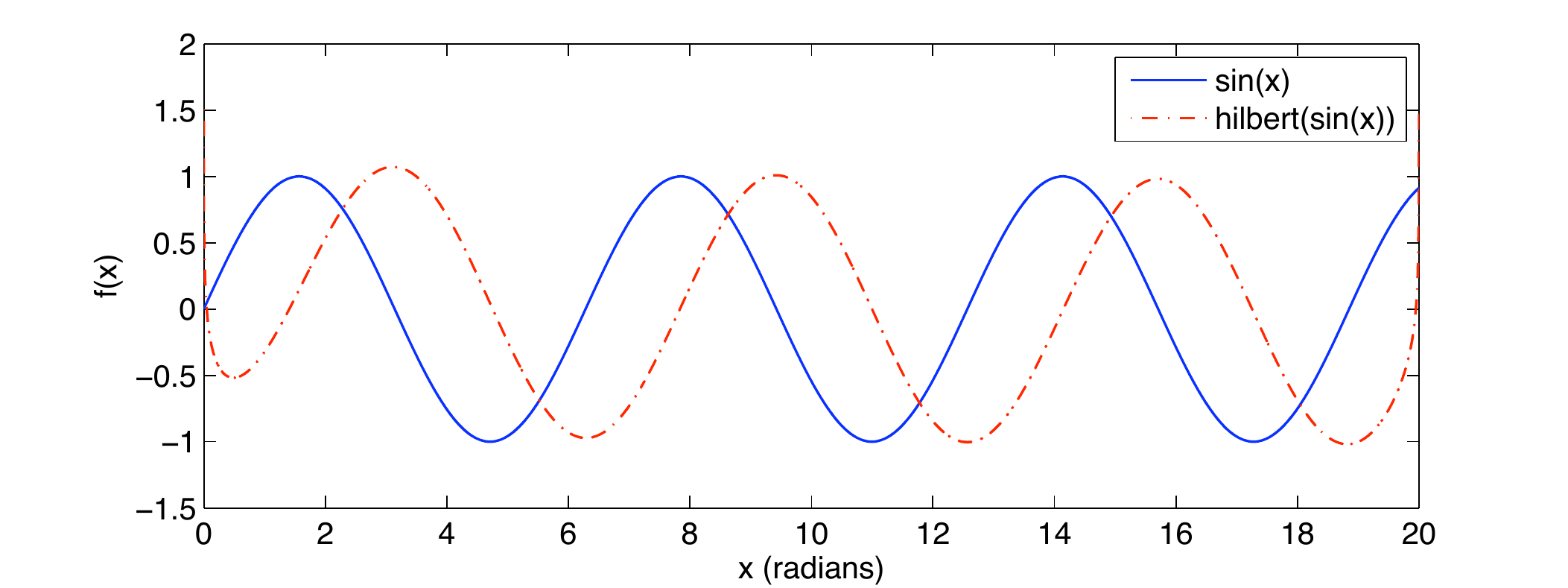}
\caption{\textbf{Example of Hilbert Transform} - The Hilbert transform causes the sine wave to become delayed by $\pi / 2$ radians which then allows for phase and amplitude information to be extracted from the original signal, through comparison of the two wave forms.}\label{fig:ExHilbert}
\end{figure}

To allow for us to analyse the phase to amplitude relationships in LFP data sets, it was first necessary for the signals to be converted to a more amenable form. This is possible by using the Hilbert transform and what is known as the analytic representation. The Hilbert transform is defined as (from \cite{Papoulis:1991}):
\begin{equation}
	y[n] = H(x[n]) = \frac{1}{\pi}\int_{-\infty}^{\infty}\frac{x[\tau]}{n-\tau}d\tau,
\end{equation}
where $y[n]$ is the transformed signal, $H$ is Hilbert transform, $x[n]$ is the original signal, and $n$ is time. When applying the Fourier transform to this result it acts as a multiplier operator. This causes all positive frequencies to be delayed by $\pi / 2$ radians. An example of the transform can be seen in Figure \ref{fig:ExHilbert} for a standard sine wave. By taking the Fourier transform this shift occurs for every frequency component of the signal. Further information about the transform can be found in \cite{LHahn:1996p9121}.

As we will be considering particular frequency bands when searching for possible coupling events, rather than using the original signal $x[n]$ we will instead use a bandpass filtered version $x_{\theta}[n]$ containing only the frequencies of interest before the Hilbert transform is applied. Using this, the analytic representation can be defined as:
\begin{equation}
	z_{\theta}[n] = x_{\theta}[n] + y_{\theta}[n]i \label{eq:analticrep}
\end{equation}
The analytic representation is a convenient way of holding the original signal, real part, and Hilbert transform, imaginary part, in a single variable. It is possible to show that this representation is equivalent to the following:
\begin{equation}
	 z_{\theta}[n] = a_{\theta}[n]\exp(\phi_{\theta}[n]i) \label{eq:phamprel}
\end{equation}
This form is useful because it allows us to extract the instantaneous amplitude $a_{\theta}[n]$ and phase $\phi_{\theta}[n]$ of the original signal at any point. These can both be written explicitly by using the ratio of original to transformed signal and Equation \ref{eq:phamprel}. Implementation of this method can be found in the \texttt{amp\_phase\_decomp.m} script and is used throughout to extract the amplitude of the high frequency carrier.
\begin{eqnarray}
	\phi_{\theta}[n] &=& \arctan\left[\frac{y_{\theta}[n]i}{x_{\theta}[n]}\right] \\ \notag \\
	a_{\theta}[n] &=& \left|\frac{x_{\theta}[n] + y_{\theta}[n]i}{\exp(\phi_{\theta}[n]i)}\right|
\end{eqnarray}

\subsection{Envelope-to-Signal (ESC) Measure}
The correlation between two variables, ranging from -1 to 1, measures any linear relationship present, giving both a strength and direction. For our purposes correlation will be defined as:
\begin{equation}
	\textrm{Corr}(x[n], y[n]) = \frac{\sum_{n=1}^{N}(x[n] - \bar{x})(y[n] - \bar{y})}{N\sigma_{x}\sigma{y}},
\end{equation}
where $x[n]$ $y[n]$ are equal length discrete input signals, $N$ is the input length, $\bar{x}$ $\bar{y}$ are signal means, and $\sigma_{x}$ $\sigma_{y}$ are standard deviations of each signal.

In order to detect any cross-frequency coupling between the phase and amplitude of two signals, correlation was used to calculate the envelope-to-signal correlation (ESC) measure defined in \cite{Bruns:2004p9011} as:
\begin{equation}
	r_{ESC} = \textrm{Corr}(x_{\theta}[n], a_{\gamma}[n])
\end{equation}
This compares how similar one frequency band varies in comparison to the change in amplitude of another. The ESC measure was chosen for this task as it had been shown in \cite{Penny:2008p8269} to perform best when detecting phase to amplitude coupling in most situations. The only issue raised was when coupling occurs during the `null phases' ($1/4$ or $3/4$ the modulating cycle) because at these points the signal is equal to its mean level, resulting in zero correlation. Unfortunately, the ESC measure cannot be adapted to counteract this problem. Instead, it would be necessary to use a measure that can detect coupling at all phases. One such measure is GLM which is presented in \cite{Penny:2008p8269}. Due to time constraints it was not possible to implement the GLM measure, however, this would make a useful extension. An implementation of the ESC measure can be found in the \texttt{calc\_corr.m} script.

\section{Modelling}

\subsection{Standard Connectionist Model}\label{sec:stdConnModel}
Before considering the effects of oscillations, a standard connectionist model was developed. This was based on the work of Usher and McClelland with neuronal groups acting as integrators for excitatory and inhibitory inputs. Each integrator is considered to be leaky, meaning the level of excitation will decrease naturally over time. A set of these integrators, each relating to a different action, are designated as outputs and a decision is made once a threshold value is exceeded by any of them. 

To incorporate the effects of noise on the system a Weiner process was used to ensure that it scales appropriately with varying time steps for numerical integration. This leads to noise being modelled as the following distribution where $dN$ is our noise output and $dt$ is the time step.
\begin{equation}
	dN \sim \textrm{Normal}\left(0,\sqrt{dt}\right)
\end{equation}

\begin{figure}[tbp]
\centering
\includegraphics[width=6.2in]{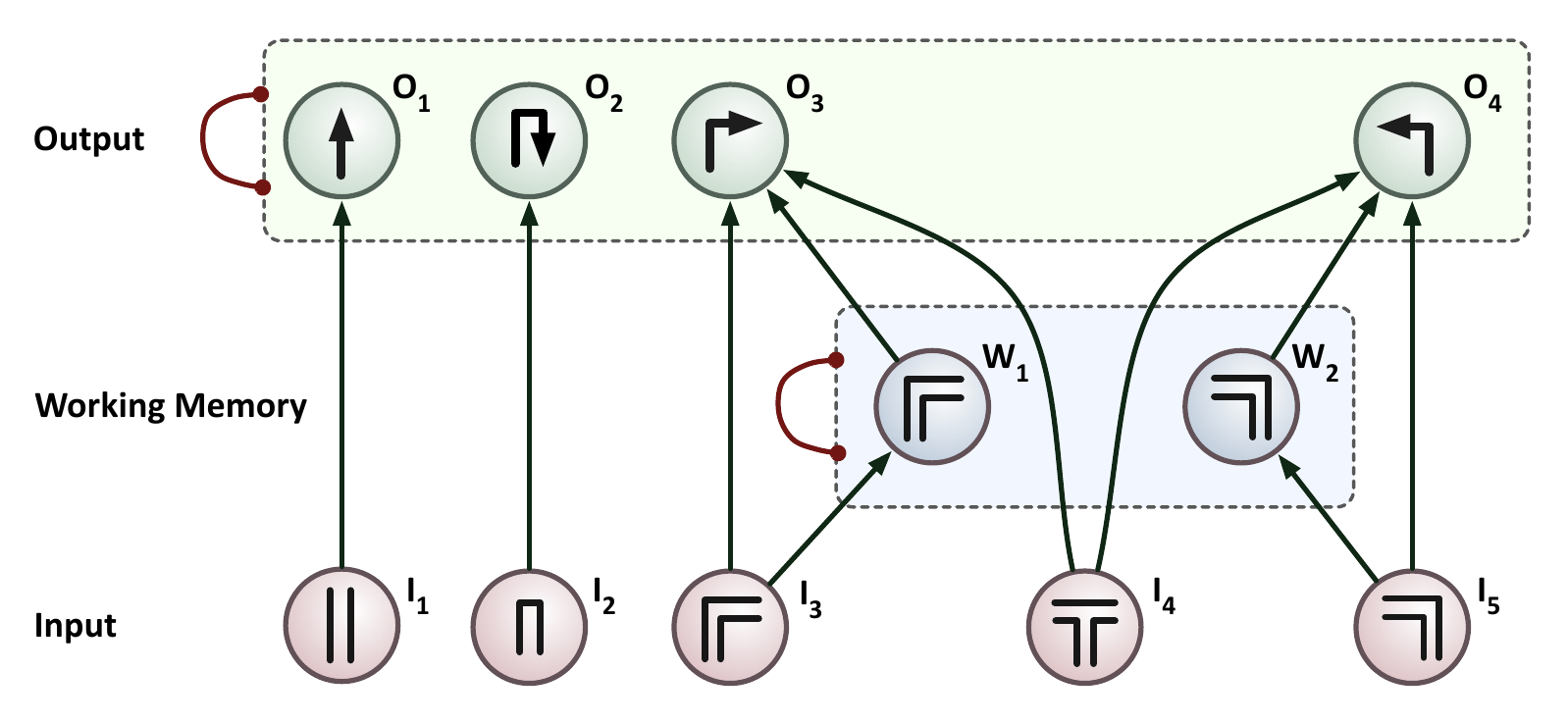}
\caption{\textbf{Standard Connectionist Model for T-Maze} - Arrows represent excitatory and dots inhibitory inputs. Each input, denoted $I_n$ represent various stimuli the rat could be presented with. For example, $I_1$ straight corridor, $I_2$ dead end, $I_3$ right hand bend, $I_4$ T-junction and $I_5$ left hand bend. The inhibition links shown on the output and working memory groups correspond to every integrator in that group having an inhibitory connections to all other integrators in the same group. This then allows for competition between integrators within the group if two are active at once.}\label{fig:ConMod_TMaze1}
\end{figure}

To carry out the decision process for the T-maze task, the model shown in Figure \ref{fig:ConMod_TMaze1} was developed. This is structured into 3 main parts -- inputs from sensory information, working memory to remember previous movements, and output defining the action that should be taken. A form of working memory is vital for this task as the subject must recall the direction of the previous forced turn in order to receive a reward. For simplicity we treat all decay constants, excitatory and inhibition rates as the same for every integrator.

The dynamics of each integrator can be defined in the following way:
\begin{equation}
	dA = \left( Ext_{1} + ... + Ext_{n} - kA - w(Inh_{1} + ... + Inh_{m}) \right) dt + cdN
\end{equation}
Here, $dA$ is the change in our integrator excitation, $Ext_{1} ... Ext_{n}$ are excitatory inputs, $k$ is the decay rate, $w$ is the inhibition rate, $Inh_{1} ... Inh_{m}$ are inhibitory inputs and $c$ is a scaling of the noise. Using this structure the output and working memory integrators can be defined as:
\begin{eqnarray}
	\dot{O_{1}} &=& (I_{1} - kO_{1} - w(O_{2} + O_{3} + O_{4}))dt + cdN \\
	\dot{O_{2}} &=& (I_{2} - kO_{2} - w(O_{1} + O_{3} + O_{4}))dt + cdN \\
	\dot{O_{3}} &=& (I_{3} + I_{4} + W_{1} - kO_{3} - w(O_{1} + O_{2} + O_{4}))dt + cdN \\
	\dot{O_{4}} &=& (I_{4} + I_{5} + W_{2} - kO_{4} - w(O_{1} + O_{2} + O_{3}))dt + cdN \\
	\dot{W_{1}} &=& (I_{3} - kW_{1} - wW_{2})dt + cdN \\
	\dot{W_{2}} &=& (I_{5} - kW_{2} - wW_{1})dt + cdN
\end{eqnarray}
Once any output integrator value exceeds the decision threshold value, the virtual rat performs the associated action. In contrast the working memory integrators do not have a threshold as they are not directly related to any actions. For all simulations a threshold value of $1$ was selected.

The simulated maze, is defined as a set of line segments designed to constrain the movement of the rat in a 2-dimensional space. Separate segments are specified between junctions and properties such as start point, end point and direction (vertical or horizontal) are held for each. The rats state is then defined as a point in 2D space, a line segment that they inhabit and a direction (up, down, left or right). This information makes it possible to set the stimuli inputs ($I_{1}, I_{2}, I_{3}, I_{4}, I_{5}$) at each time step by effectively matching what the rat sees at its present location to one of the possible stimuli. These are then set to $1$ when active and $0$ otherwise.

Although we consider a discrete set of possible directions, the rat's location in the 2D environment is updated in a continuous manner. When an output integrator reaches the threshold, the rat's location is updated based on the current direction and a user defined speed. Constraints in terms of the rat attempting to move outside a maze segment are also taken into consideration, and if an attempt is made to move through a wall, the rat's position remains fixed.

\begin{figure}[tbp]
\centering
\includegraphics[width=3.8in]{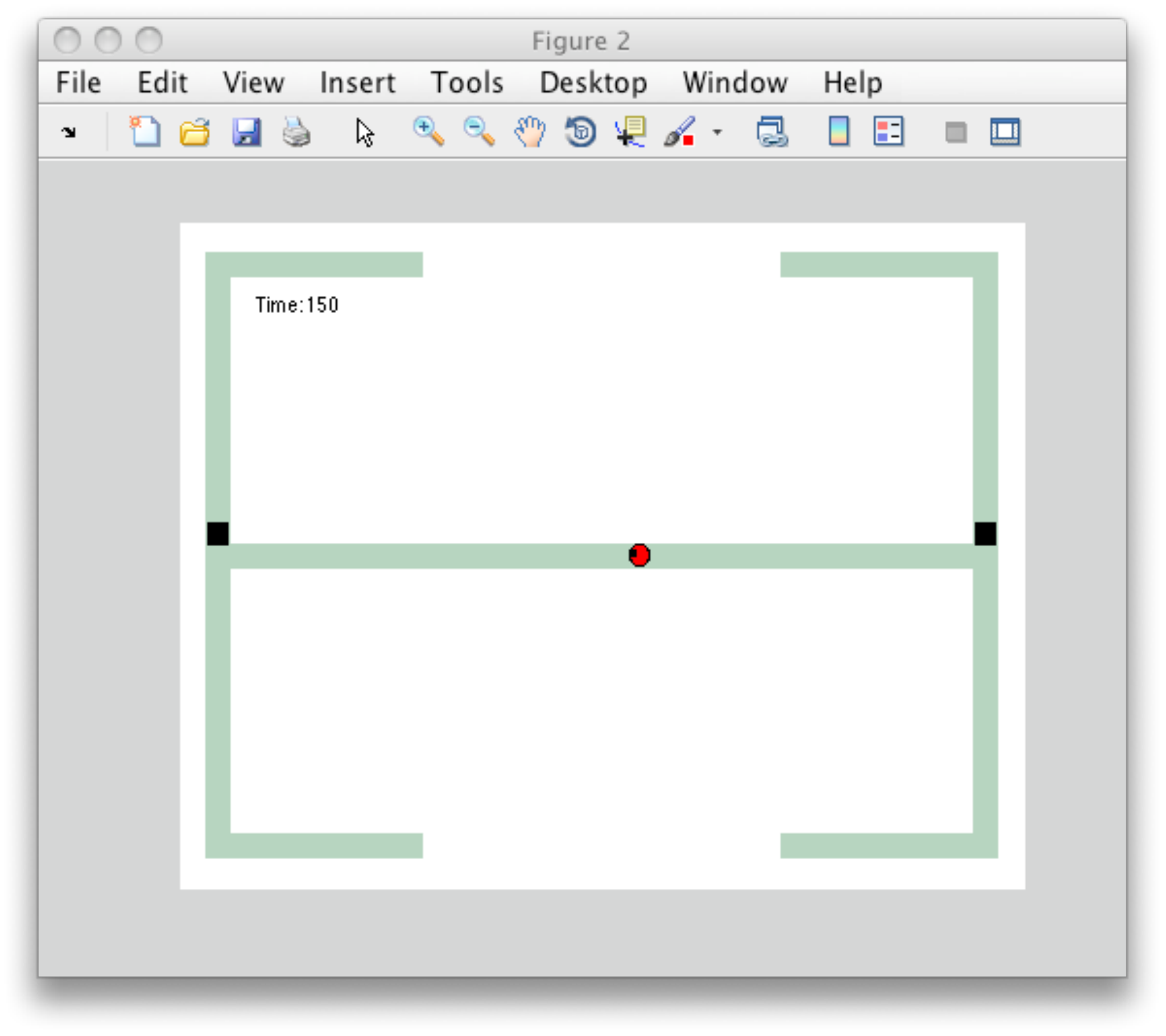}
\caption{\textbf{Visualisation of Standard T-Maze Model} - To allow for behavioural aspects of the rats movement in the model to be assessed an animation of the modelled rats movement is available. Here, the maze is shown in green, blocks in black and the rat in red with a small black dot specifying the head direction.}\label{fig:ratSimVisual}
\end{figure}

This model was implemented in the \texttt{rat\_model\_cont.m} script and parameters were selected through experimentation with the simulation. It is possible for this script to also generate an animated visualisation, allowing for behavioural aspects of the rat's movement to be analysed. An example of this output is shown in Figure \ref{fig:ratSimVisual}.

\subsection{Model Including Oscillations}
To incorporate oscillations into the existing model we chose to alter the way inhibition was configured within the working memory and output groups. Previously, we had merely included inhibitory links to every other integrator. This allowed for conflict resolution, however, in the actual brain there is evidence that a shared group of inhibitory neurones is used. This required alteration of the model to include an additional two integrator units specifically for inhibition in the output and working memory groups.

\begin{figure}[tbp]
\centering
\includegraphics[width=5.5in]{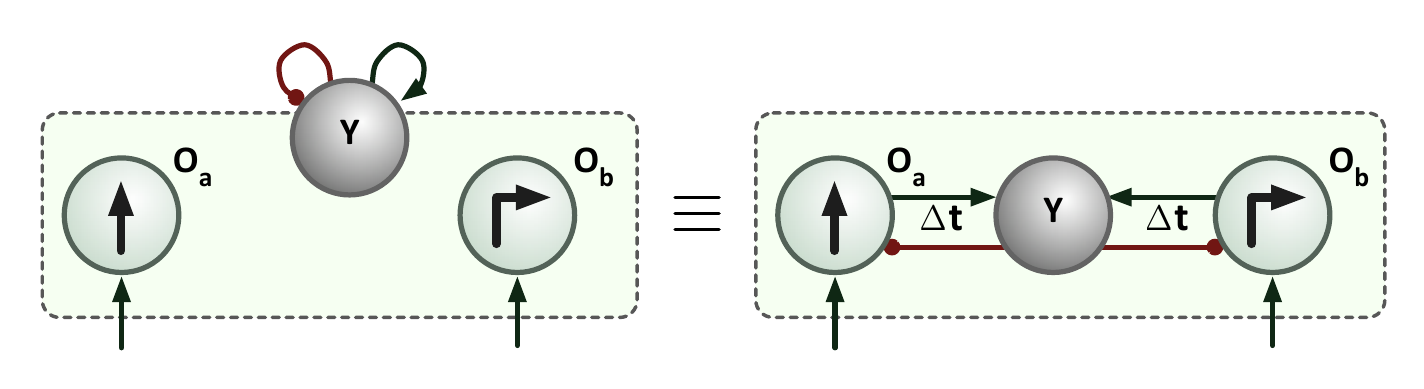}
\caption{\textbf{Delayed Excitation and Inhibition} - For simplicity the left hand diagram will be used as equivalent to the model shown on the right hand side. By updating each of the integrator groups to this form, with the delay $\Delta t$ for excitation and inhibition, it is possible to show that oscillations will naturally form.}\label{fig:ConMod_SharedInh}
\end{figure}

Adding these alone would not lead to the formation of oscillations as they effectively reproduce the existing model, using additional connections. Instead, it is possible to show that by considering a delay $\Delta t$ in the connections to and from the shared inhibitory units, oscillations will naturally arise when $\Delta t$ is greater than $0$. This type of configuration is shown in Figure \ref{fig:ConMod_SharedInh}.

\begin{figure}[tbp]
\centering
\includegraphics[width=6.3in]{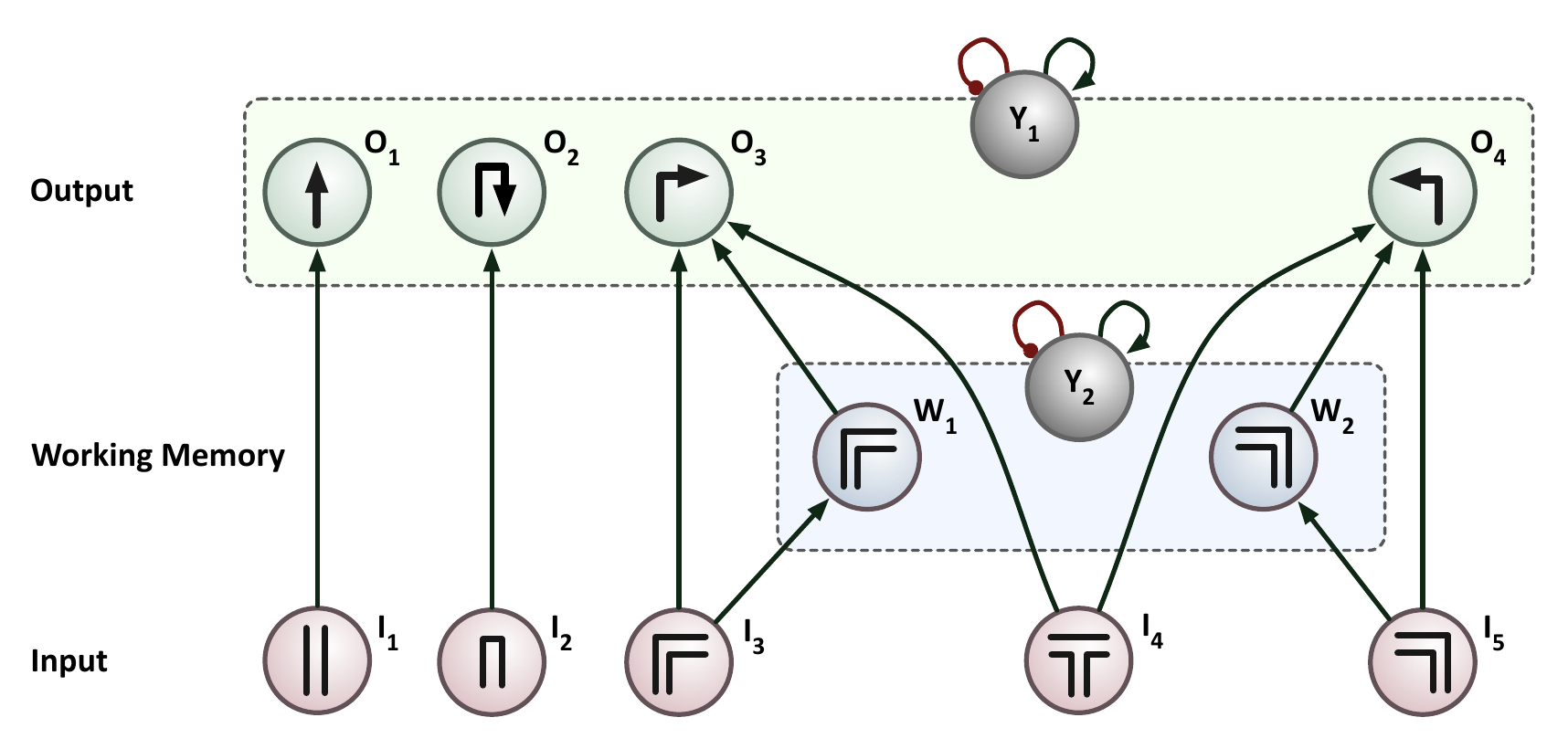}
\caption{\textbf{Oscillating Connectionist Model for T-Maze} - The original connectionist model Figure \ref{fig:ConMod_TMaze1} has been updated to incorporate a shared inhibitory unit for each group of integrators and delays in excitation and inhibition in each of these groups.}\label{fig:ConMod_Osc}
\end{figure}

Although it is possible to consider delays in all the connections, for our model they were only included to and from the shared inhibitory units. The reason for this choice is to minimise changes to the original model and assess the fundamental aspects that lead to oscillations. Due to the shear number of connections that would be required to display the whole model incorporating these changes, a simplified notation, see Figure \ref{fig:ConMod_SharedInh}, was used in the final model shown in Figure \ref{fig:ConMod_Osc}.

To mathematically describe this model the previous template, see Section \ref{sec:stdConnModel}, for each integrator was used. In addition, two constant factors were added, $a_1$ and $a_2$ to allow for the shared inhibitory units excitation to be scaled due to the large number of connections each receives. The final equations describing the oscillating model are shown below with $t_d = t - \Delta t$. Implementation of this model in MATLAB can be found in the \texttt{rat\_model\_osc.m} script. 

\begin{eqnarray}
	\dot{O_1}(t) &=& [I_1(t) - kO_1(t) - wY_1(t_d)]dt + cdN \\
	\dot{O_2}(t) &=& [I_2(t) - kO_2(t) - wY_2(t_d)]dt + cdN \\
	\dot{O_3}(t) &=& [I_3(t) + W_1(t) + I_4(t) - kO_3(t) - wY_1(_d)]dt + cdN \\
	\dot{O_4}(t) &=& [I_4(t) + W_2(t) + I_5(t) - kO_4(t) - wY_1(_d)]dt + cdN \\
	\dot{W_1}(t) &=& [I_3(t) - kW_1(t) - wY_2(t_d)]dt + cdN \\
	\dot{W_2}(t) &=& [I_5(t) - kW_2(t) - wY_2(t_d)]dt + cdN \\
	\dot{Y_1}(t) &=& [a_1[O_1(t_d) + O_2(t_d) + O_3(t_d) + O_4(t_d)] - kY_1(t)]dt + cdN \\
	\dot{Y_2}(t) &=& [a_2[W_1(t_d) + W_2(t_d)] - kY_2(t)]dt + cdN
\end{eqnarray}

\cleardoublepage
\chapter{Results}

\section{Cross Frequency Coupling}
Several tools were developed in MATLAB to help search for cross-frequency coupling between phase and amplitude. To test that these were giving accurate results, analysis was first performed on generated data with known relationships being present. 

\begin{figure}[tbp]
\centering
\includegraphics[width=6.3in]{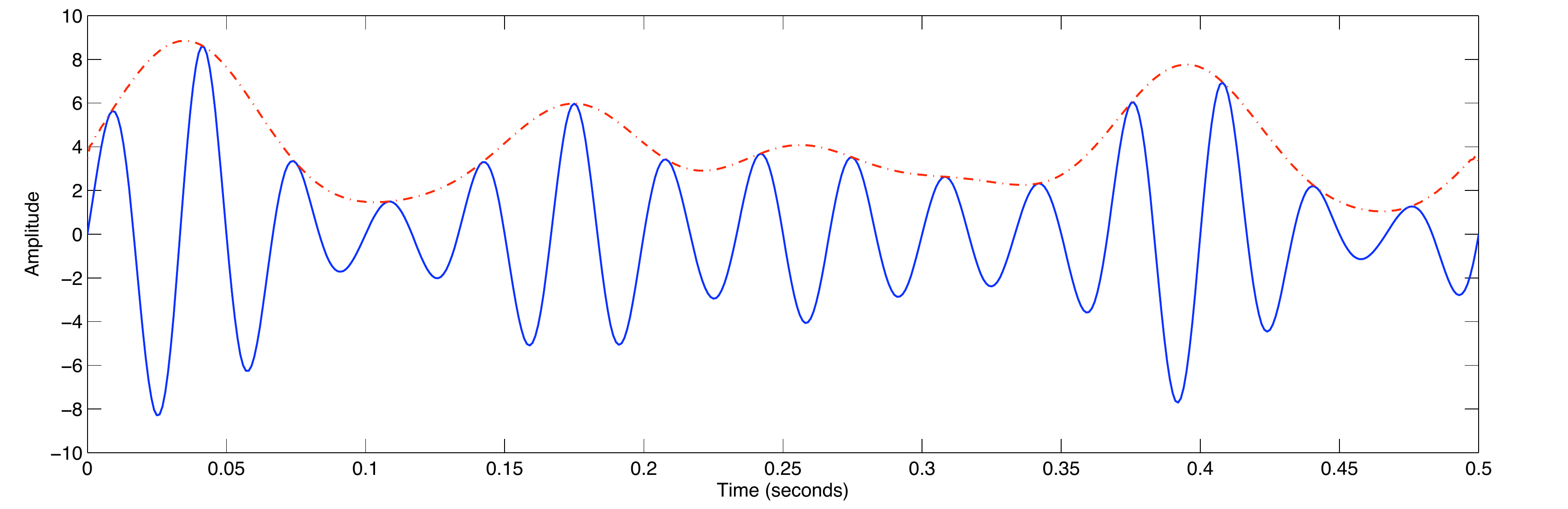}
\caption{\textbf{Amplitude Decomposition} - Blue line represents the input signal and the red dotted line the estimated amplitude, found using the method described in Section \ref{sec:ampDec}.}\label{fig:PowPhaCouple2}
\end{figure}

The first aspect to test was the amplitude decomposition (\texttt{amp\_phase\_decomp.m}) which was used throughout other analysis techniques. Figure \ref{fig:PowPhaCouple2} shows an example of a result from testing. The solid blue line represents a custom modulated signal, while the dotted red line is the estimated amplitude reconstruction. This illustrates that the method is working accurately and gave us confidence for its further use. This technique is reliant on the underlying frequencies being modulated, with the reconstruction estimate having a higher accuracy for carrier signals with a higher frequency. Previous work \cite{Tort:2008} showed that theta (4-12Hz) modulates gamma (30-90Hz) which meets the theoretical Nyquist rate for reconstruction of the lower frequency signal without aliasing. It is worth understanding that if the Nyquist rate is not met then accurate reconstruction of a low frequency modulating phase may not be possible and therefore would be an unlikely candidate for use in the brain.

\begin{figure}[tbp]
\centering
\includegraphics[width=4.7in]{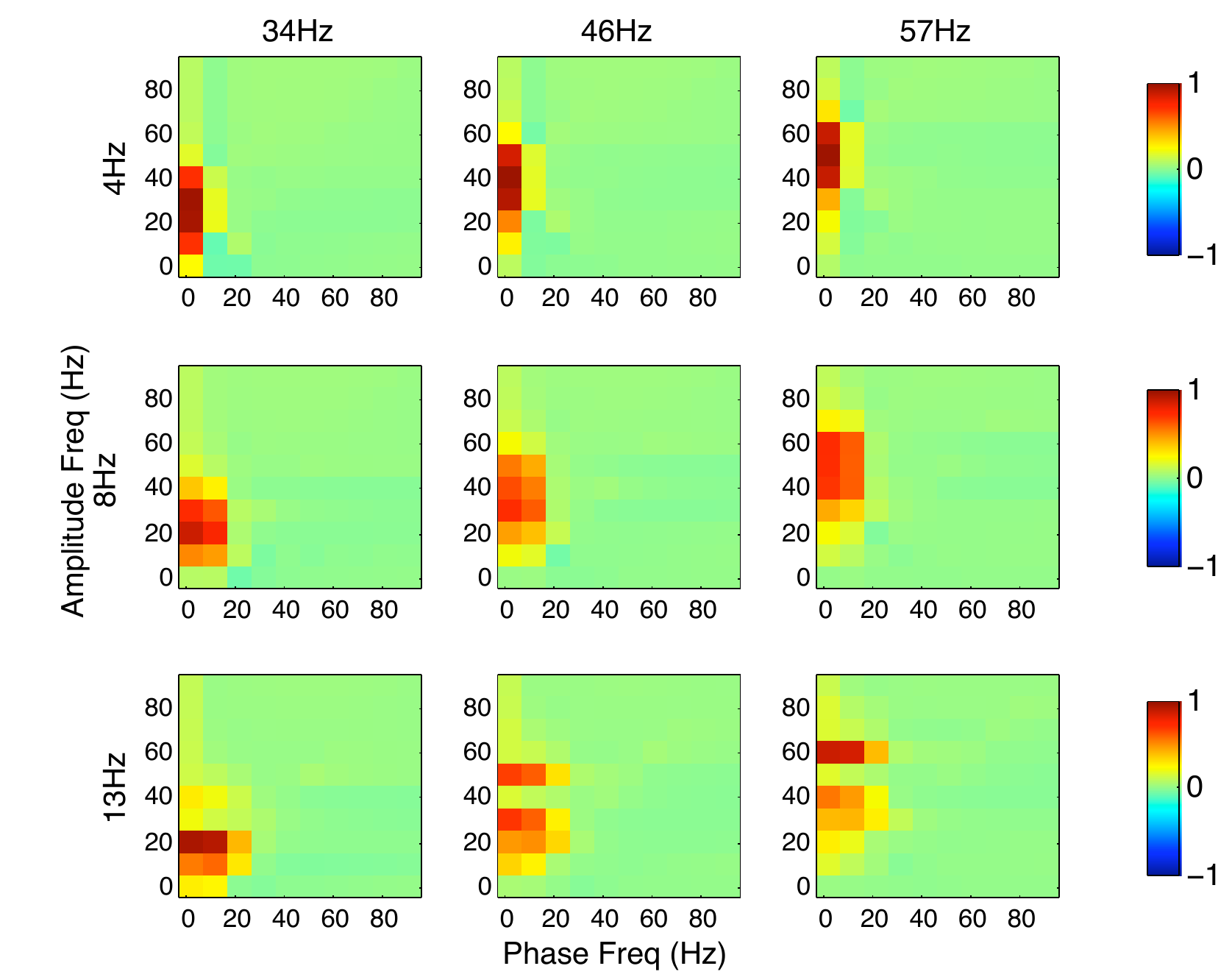}
\caption{\textbf{Testing ESC Correlation Measure} - Test data was generated with differing phase and amplitude coupling properties. Moving to the right, plots have an increased amplitude frequencies of 34Hz, 46Hz and 57Hz, while moving down the plots leads to increased phase frequencies of 4Hz, 8Hz and 13Hz. \\ \vspace{-1.2cm}}\label{fig:CoupTest}
\end{figure}

Once the method for amplitude decomposition had been shown to work, we were able to test for coupling via the ESC measure. Again to ensure this was working correctly, test data with known relationships was generated and correlation visualised as a 2D heat plot. Figure \ref{fig:CoupTest} shows the results from this testing. As you move across to the right the high frequency amplitude is increased resulting in the correlation point moving upwards. Moving down the plots results in the low frequency phase being increased causing the correlation to move towards the right. We only concentrated on phase frequencies in the 4-13Hz range as these were of interest to us and could accurately be reconstructed using carrier frequencies less than 100Hz. All filtering was performed using Butterworth filters of order $2$ and designed using the \texttt{butter} function. This filter was then applied both in forward and reverse directions to ensure zero-phase shift in the output signal. The filtering process was implemented using the standard \texttt{filtfilt} function.

Test results show that in general high areas of correlation relate to the underlying relationships present. Lower frequency phases are found with greatest accuracy which may be due to the limited accuracy possible in reconstructing the amplitude with 34-57Hz carrier signals.

During testing it was evident that many factors have a large impact on the results that are obtained. Originally it was hoped that the process of detecting coupling relationships could be automated based on any two input recordings. This is not possible, however, due to the suitable differences between recordings and the possible types of coupling that may occur. For example, if the coupling occurs over a long period of time it is possible to use filters of larger order and larger windows for analysis. This gives much smoother results and helps to reduce artefacts present. Conversely, if events occur over a fairly short period, which seems to be the case for this type of coupling, it is necessary to use shorter filters and analysis windows to enable a high time resolution. This, however, leads to artefacts as not all frequencies can be properly described in the signal lengths used.

\begin{figure}[tbp!]
\centering
\subfloat[\textbf{Forced Turn}]{
\includegraphics[width=5.2in]{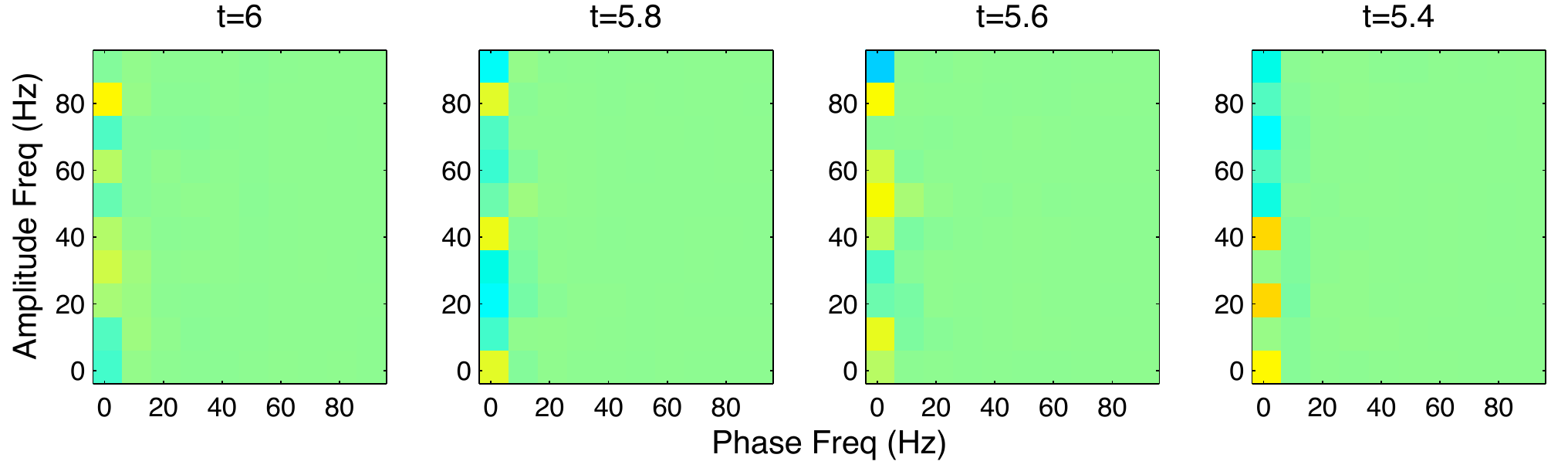}
\includegraphics[width=0.45in]{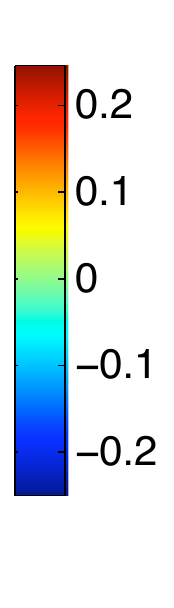}}\\
\subfloat[\textbf{Choice Turn}]{
\includegraphics[width=5.2in]{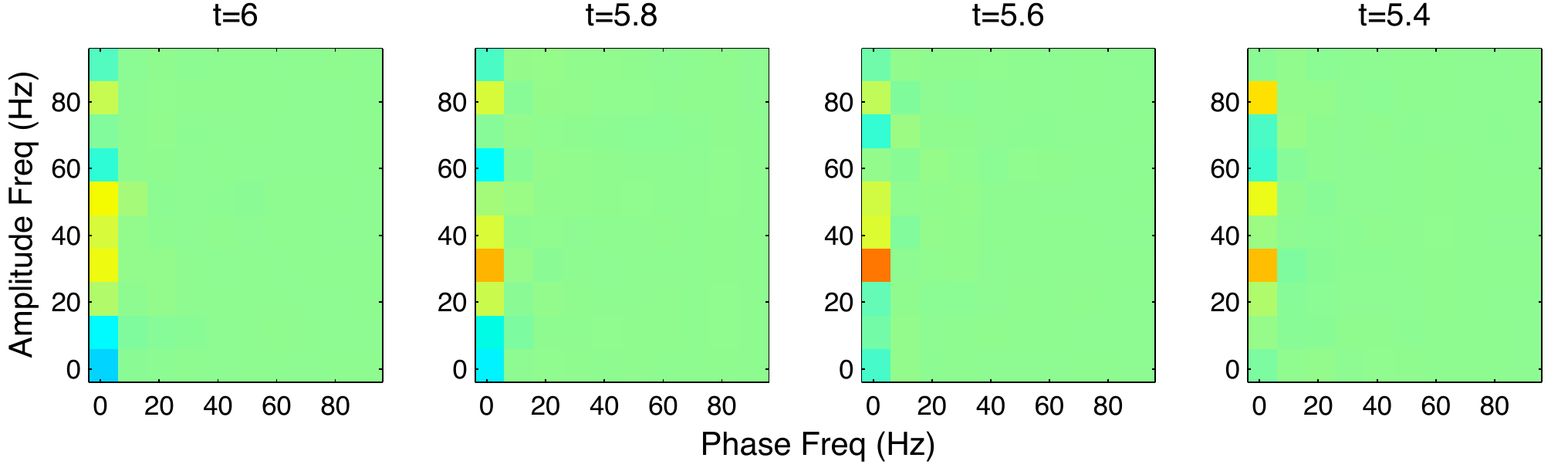}
\includegraphics[width=0.45in]{DS2_Ts_Corr_scale.pdf}}
\caption{\textbf{ESC Measure Approaching Forced and Choice Turns} - Each sub plot shows ESC measure for the phase of HPC to the amplitude frequency of PFC in the range 1-100Hz. Results were generated from the `Motorhead day d' data set. $t=6$ is the point where the choice or forced turn is made and final correlation values were calculated as an average of 16 trails. It is possible to see increased correlation for the choice turn, with the HPC 1-10Hz theta band modulating the PFC 30-40Hz low gamma band.}\label{fig:DS2_Ana}
\end{figure}

Having shown that the techniques worked with derived data, analysis was next carried out on an actual LFP recordings. The data set chosen was for the rat named `Motorhead' and recordings from day d. This data is known to include increased coherence between the PFC and HPC signals when reaching a choice turn and was seen as a good starting point to see how well the techniques worked with real data containing noise. The data set comprises of 16 trials and is split into separate files relating to a 6 second period up to choice or forced turns. Previous analysis had shown that correlation increased close to the choice turn and so we focused on the last 0.6 seconds of the recording. Results of the coupling analysis are shown in Figure \ref{fig:DS2_Ana}. These plots average the ESC measure for the 16 trails at 4 time points leading up to the choice or forced turn. From these the forced turn results show areas of increased correlation for short periods of time but at relatively low levels. Interestingly, all correlations that do occur have the 1-10Hz theta band as the phase frequency for modulation.

The choice turn results also show sporadic correlations in the theta band, however, a persistent increased correlation of up to 0.2 is seen between the 30-40Hz low gamma band and 1-10Hz theta band. This appears to peak at around 0.4 seconds before the choice turn is made. Although this correlation value is still fairly low, the persistent nature of the event supports the possibility that significant coupling is occurring.

\begin{figure}[tbp!]
\centering
\subfloat[\textbf{Forced Turn}]{
\includegraphics[width=5.2in]{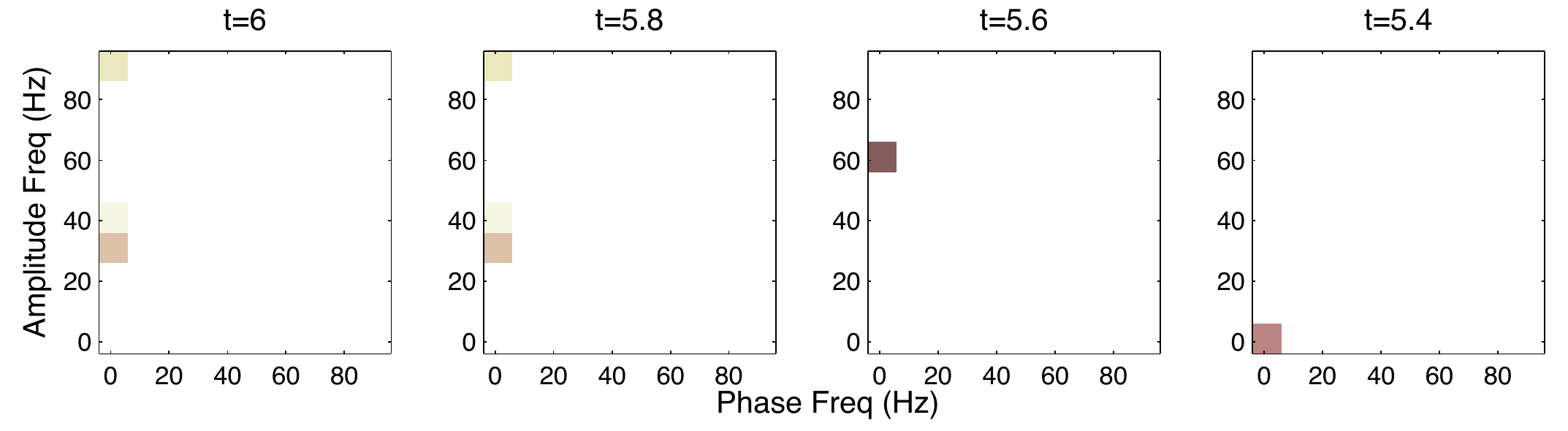}
\includegraphics[width=0.65in]{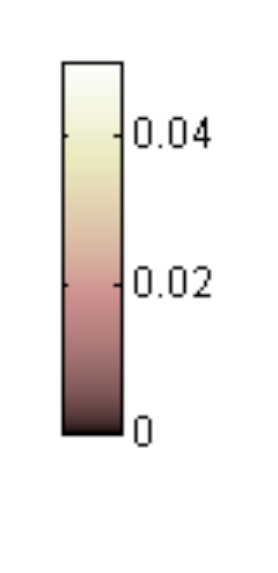}}\\
\subfloat[\textbf{Choice Turn}]{
\includegraphics[width=5.2in]{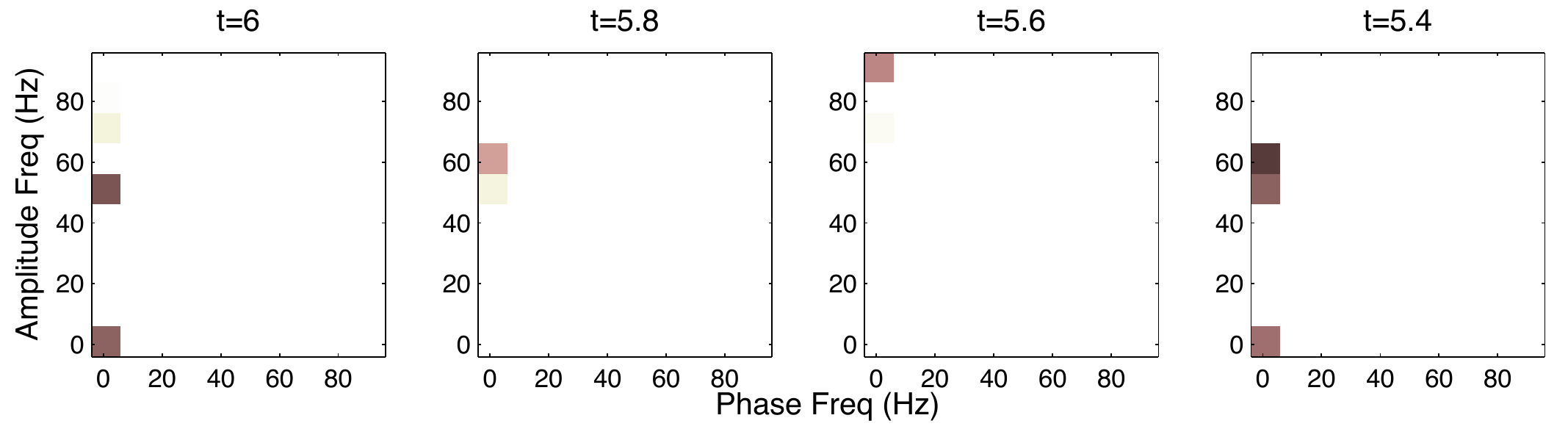}
\includegraphics[width=0.65in]{CorrTestScale.pdf}}
\caption{\textbf{Correlation Significance Analysis} - Each sub plot displays the average p-value for the results shown in Figure \ref{fig:DS2_Ana}. Only significant results with a p-value < 0.05 are displayed, larger values are ignored and shown as white.}\label{fig:DS2_Corr}
\end{figure}

In an attempt to gain insight into the statistical significance of these correlations, average p-values were calculated for each combination of frequencies. These results are shown in Figure \ref{fig:DS2_Corr}. As we are only interested in significant correlations, the plots only highlight areas where the p-value is less than 0.05. It was hoped that the persistent correlation between the 30-40Hz low gamma band and 1-10Hz theta band during a choice turn would be shown to be significant, however, it was not. Instead, the only real persistent significance seem to be for the 50-70Hz gamma band and 1-10Hz theta band during choice turns.

One of the possible reasons why no significant correlation was found where the coupling was greatest may have been due to averages being taken of the p-values. To see if this was the case, p-values for each trial were calculated. This showed that for the frequencies we had highlighted with persistent high coupling, 11 out of 16 had a significant correlation. Due to a few p-values being very large (0.6664, 0.3908), the average result had become insignificant, even though in the majority of cases the value was very small. Obviously, we would want every trail to show a significant result for these frequencies, however, it is likely that variances in noise and the exact time periods over which the data is recorded will all effect how significant the correlation appears. With the large majority of trails showing a significant correlation between the coupling of the 30-40Hz low gamma band and 1-10Hz theta band during a choice turn, this gives us some confidence in pursuing the area with more advanced analysis techniques in the future.

\section{Standard Connectionist Model}
\begin{figure}[tbp]
\centering
\subfloat[\textbf{Output Integrators}]{
\includegraphics[width=6.4in]{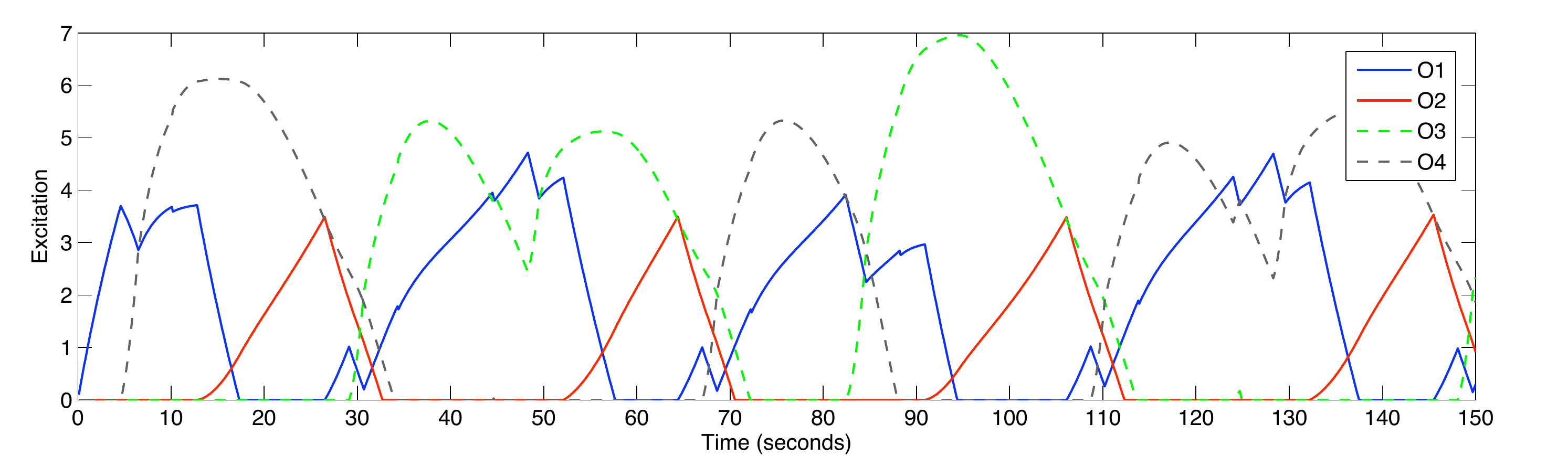}}\\
\subfloat[\textbf{Working Memory}]{
\includegraphics[width=6.4in]{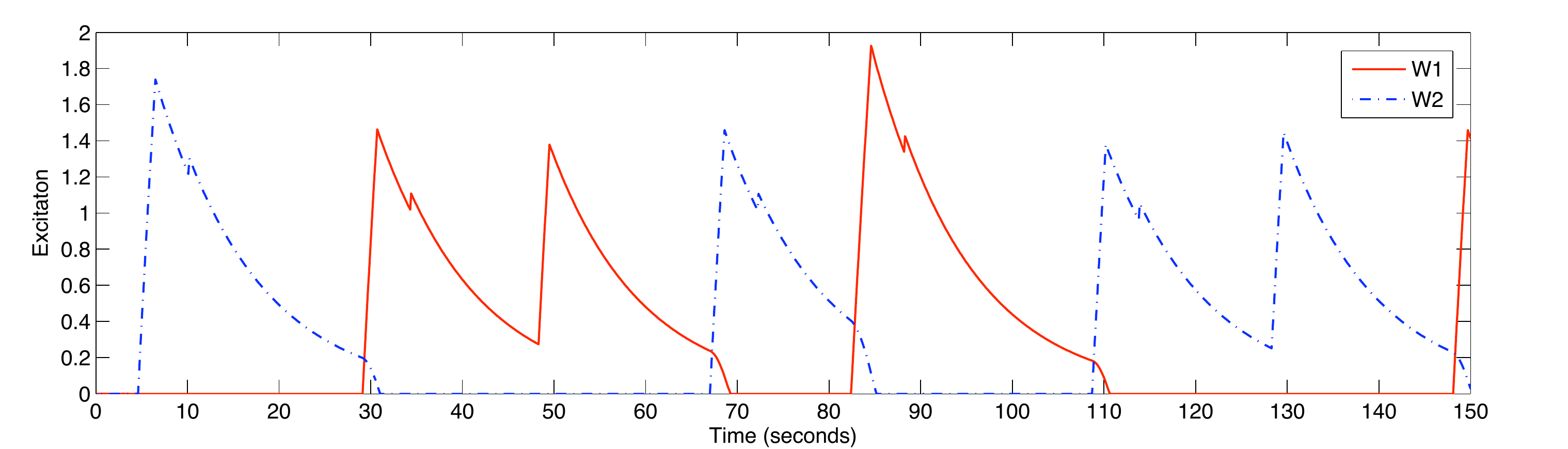}}
\caption{\textbf{Standard Model Dynamics} - The dynamics of the standard model with parameter values of $dt=0.1$, $w=0.2$, $k=0.2$, $c=0.0001$ and a decision threshold of 1.\\ \vspace{-1.2cm}
}\label{fig:ResStdModelStates}
\end{figure}

Rather than attempt to incorporate oscillations into a model immediately, a standard model was developed first to act as a baseline on which to compare others. Parameters were selected by experimentation with the simulation and attempting to approximately match experimental results (success rate $\approx$ 90\% with standard deviation $\approx$ 8\%). This lead to the following parameters being chosen, $dt=0.1$, $w=0.2$, $k=0.2$ and $c=0.0001$ which gave a simulated success rate of 87.9\% with standard deviation of 9.1\%. Results from this model are shown in Figure \ref{fig:ResStdModelStates} with output and working memory integrator groups plotted separately. From these plots the periodic behaviour of the integrators is evident. This is partly an artefact of the maze design as the lengths of sections are relatively similar, causing stimuli to be active at set periods.

The working memory shows that only an individual integrator is active at any one time based on the previous forced turn that has been seen. This also shows multiple peaks on some of the decay periods, these are due to there being two forced turns required to exit an arm and return to the joining section of the maze. The period of the decay is important because if too rapid, when the choice turn is reached there will be no bias in the decision and effectively a random choice will be made. It can be seen that neither of the working memory integrators reach zero before a choice turn is made. 

The output integrators also have a periodic nature depending on which portion of the T-maze the rat is located. The decay of excitation is much steeper than for the working memory, this is due to there always being one stimuli active, and the inhibition within the group  effectively increasing the factor by which the integrator is inhibited.

\section{Modelling of Oscillations}
Having produced a baseline model, oscillations were now added by using shared inhibitory units and incorporating a delay in all connections to and from these. These changes introduced 3 additional parameters, the two factors $a_1$, $a_2$ for shared inhibitory units excitation and a delay $\Delta t$. Again through experimentation $\Delta t$ was selected to be 0.1 while varying $a_1$ and $a_2$ lead to some interesting dynamics.

\begin{figure}[p!t]
\centering
\subfloat[\textbf{Output Integrators}]{
\includegraphics[width=6.4in]{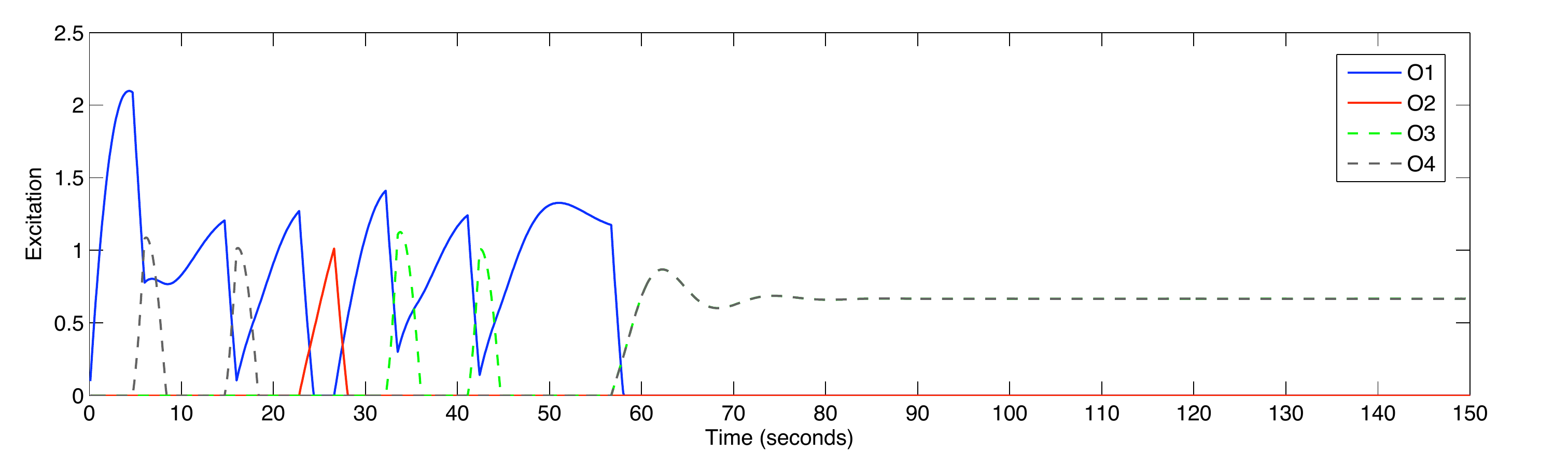}}\\
\subfloat[\textbf{Working Memory}]{
\includegraphics[width=6.4in]{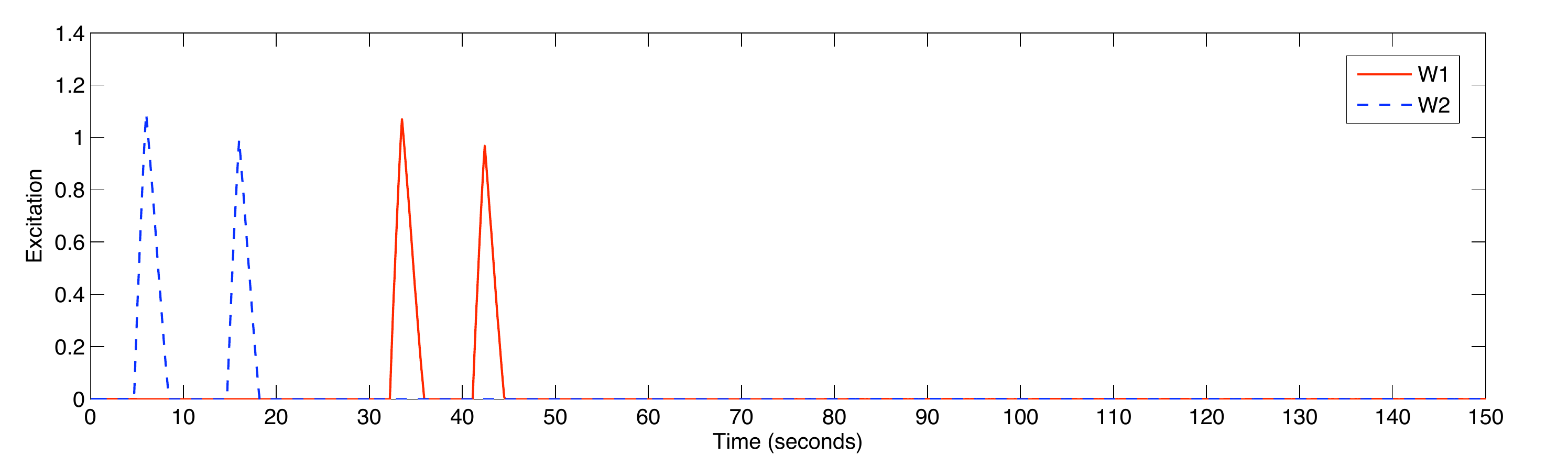}}\\
\subfloat[\textbf{Shared Inhibitory Units}]{
\includegraphics[width=6.4in]{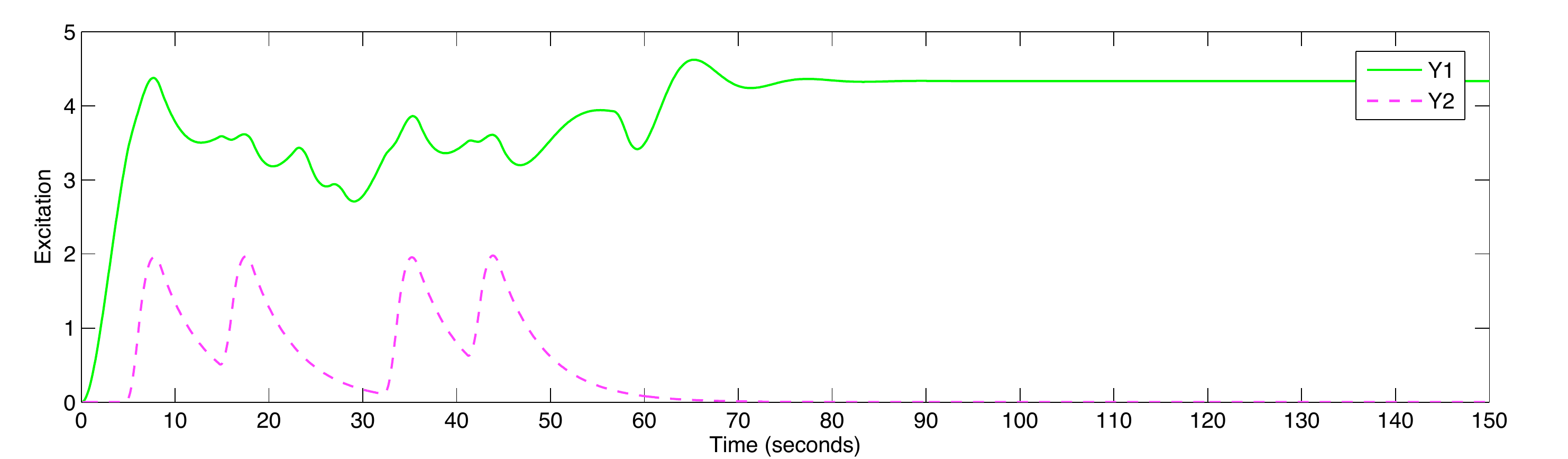}}
\caption{\textbf{Oscillating Model Dynamics Reaching Stable State} - The dynamics of the oscillating model using shared inhibition rates of $a_1 = 1$ and $a_2 = 2$. Other parameters were set as: $dt = 0.1$, $w=0.2$, $k=0.2$, $c=0.0001$ and $\Delta t = 0.1$. With this simulation the inhibition of the shared integrators is very high, shown by the steep decay for both output and working memory integrators. The quick decay of working memory causes no major bias in left and right outputs ($O_3$, $O_4$) at the choice turn ($t\approx 60$) leading to the excitation of the shared inhibitor reaching a level which drives both left and right outputs below the required decision threshold of 1. \\ ~ \\ ~ \\}\label{fig:ResOscModelStates_Stable}
\end{figure}

When $a_1$ and $a_2$ was set to values $>1$ the simulation would reach a steady state where the rat would halt unable to move. Results from this type of configuration are shown in Figure \ref{fig:ResOscModelStates_Stable}. This is due the output integrators reaching a constant state below the decision threshold of 1. The reason this dynamic arises is due to the inhibitory units driving the working memory down to zero before the choice turn is made. This causes both left and right turn outputs to rise, however, this results in the shared inhibitory unit excitation increasing quicker to a level that stops the outputs ever reaching the threshold. Obviously such a dynamic in the real-world would be highly detrimental and so it is unlikely that such a situation could ever be reached.

\begin{figure}[p!t]
\centering
\subfloat[\textbf{Output Integrators}]{
\includegraphics[width=6.4in]{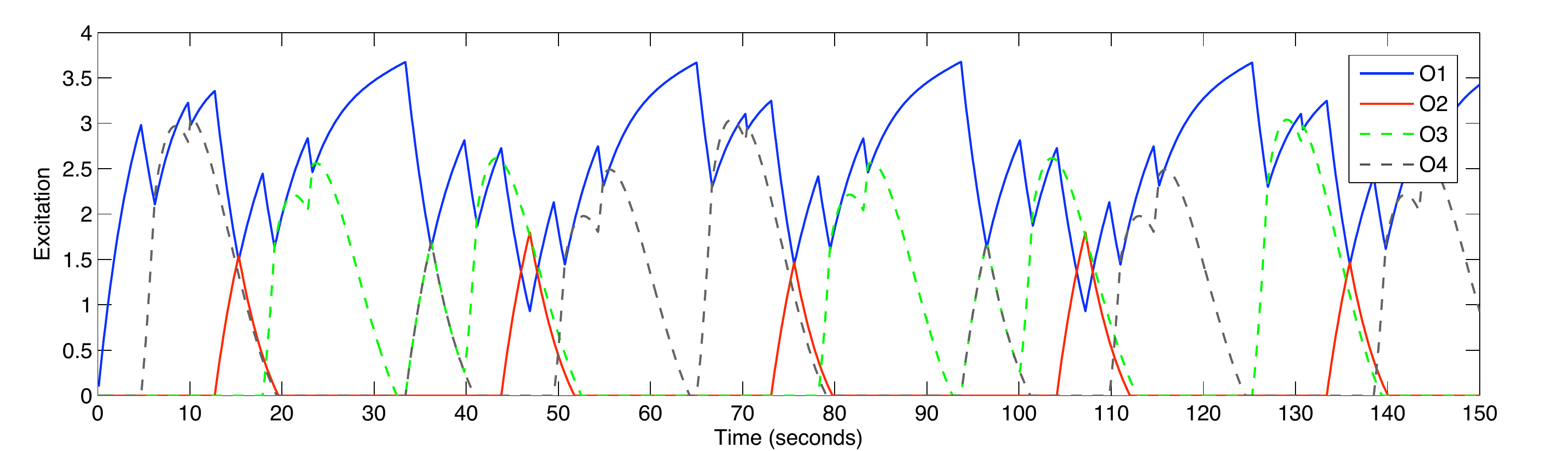}}\\
\subfloat[\textbf{Working Memory}]{
\includegraphics[width=6.4in]{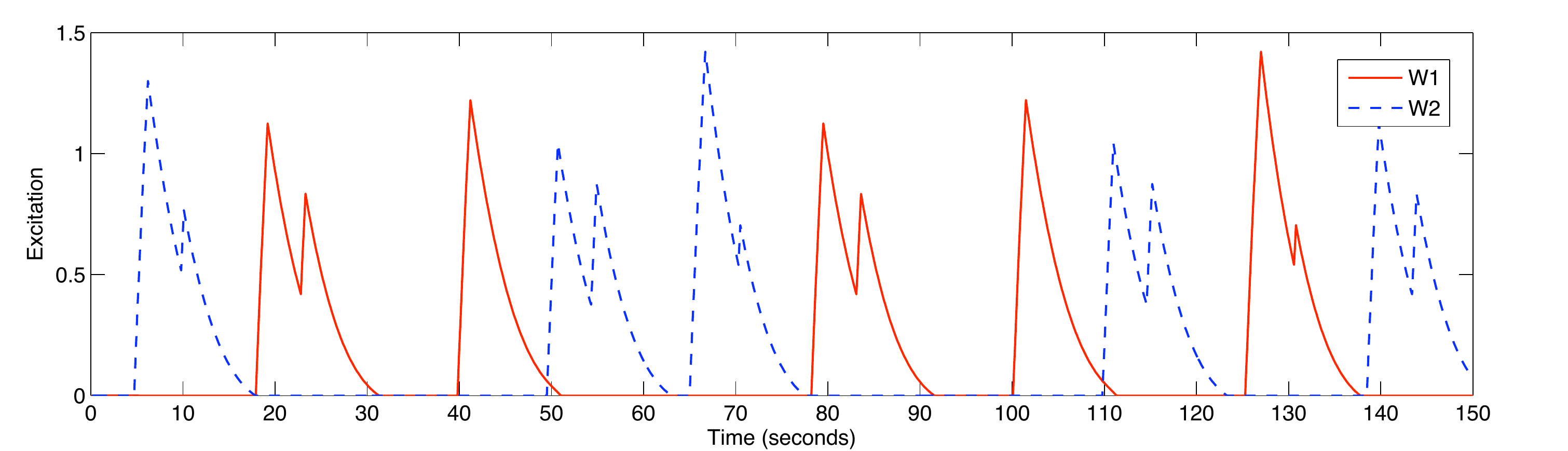}}\\
\subfloat[\textbf{Shared Inhibitory Units}]{
\includegraphics[width=6.4in]{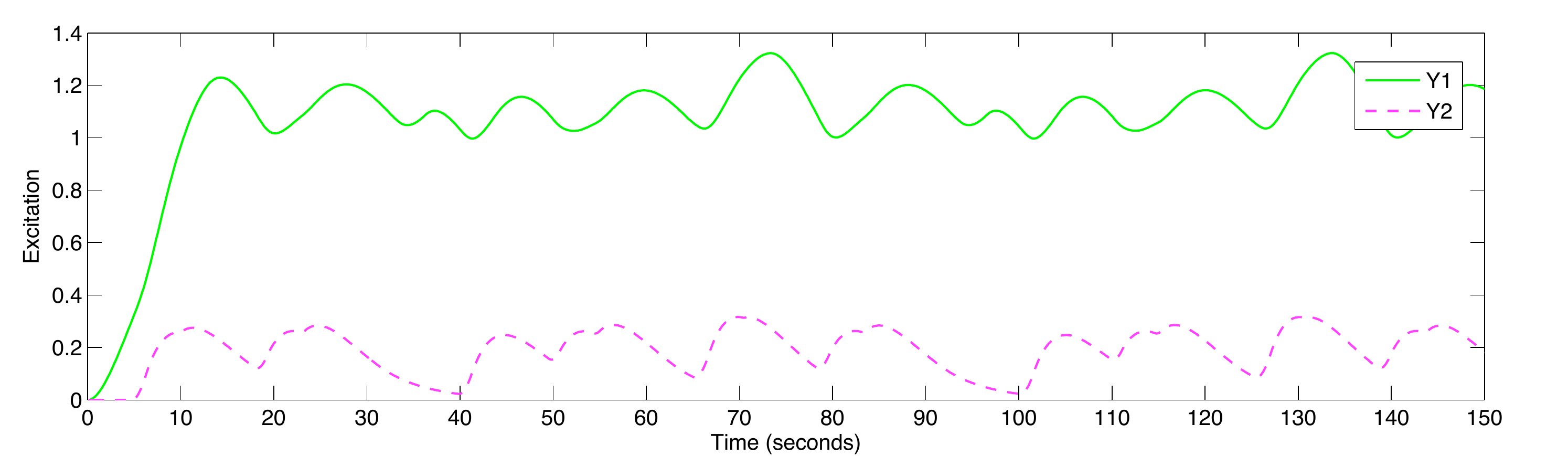}}
\caption{\textbf{Unstable Oscillating Model Dynamics} - The dynamics of the oscillating model using shared inhibition rates of $a_1 = 0.05$ and $a_2 = 0.1$. Other parameters were set as: $dt = 0.1$, $w=0.2$, $k=0.2$, $c=0.0001$ and $\Delta t = 0.1$. This simulation uses a smaller decay rate for the shared inhibitors, causing working memory to decay more slowly and allowing it to influence the choice turn. This additional contribution always allows one of the output integrators to reach the threshold and removing the chance of a stable equilibrium being reached. Although oscillations are not present in the output and working memory integrators, a regular rhythm is displayed by the shared inhibitory units. \\ \vspace{1.2cm} }\label{fig:ResOscModelStates_Unstable}
\end{figure}

The alternative dynamic seen was when $a_1$ and $a_2$ were set to much smaller values with $0.05$ and $0.1$ working well. Output from a simulation using these settings is shown in Figure \ref{fig:ResOscModelStates_Unstable}. Here it is possible to see that the inhibition rate is much less allowing the working memory integrators to decay more slowly. This in turn allows for the output integrators to always reach the threshold and for a decision to be made. An interesting aspect of this simulation is the excitation of the shared inhibitory units. Both appear to oscillate for the duration of the simulation. It can be shown that mathematically this result is valid, however, it is interesting to attempt and relate this to real-world recordings. As stated previously, LFP recordings are thought to measure the synchronised input into a region, not the output. The shared inhibitory units are connected, indirectly, to all input stimuli and it may be that LFP recordings are more akin to displaying their output. 

Due to time constraints it was not possible to cover a large number of parameter values. This means that although oscillations were only seen in the shared inhibitory units, it may be possible for the system to exhibit all integrators oscillating. This would be necessary if they are to be used as a form of co-ordination between regions, unless the shared inhibitory units have connections between groups of integrators.

\section{The Effects of Noise}
\begin{figure}[pb]
\centering
\subfloat[\textbf{Average \% of Successful Trails}]{
\includegraphics[width=5.6in]{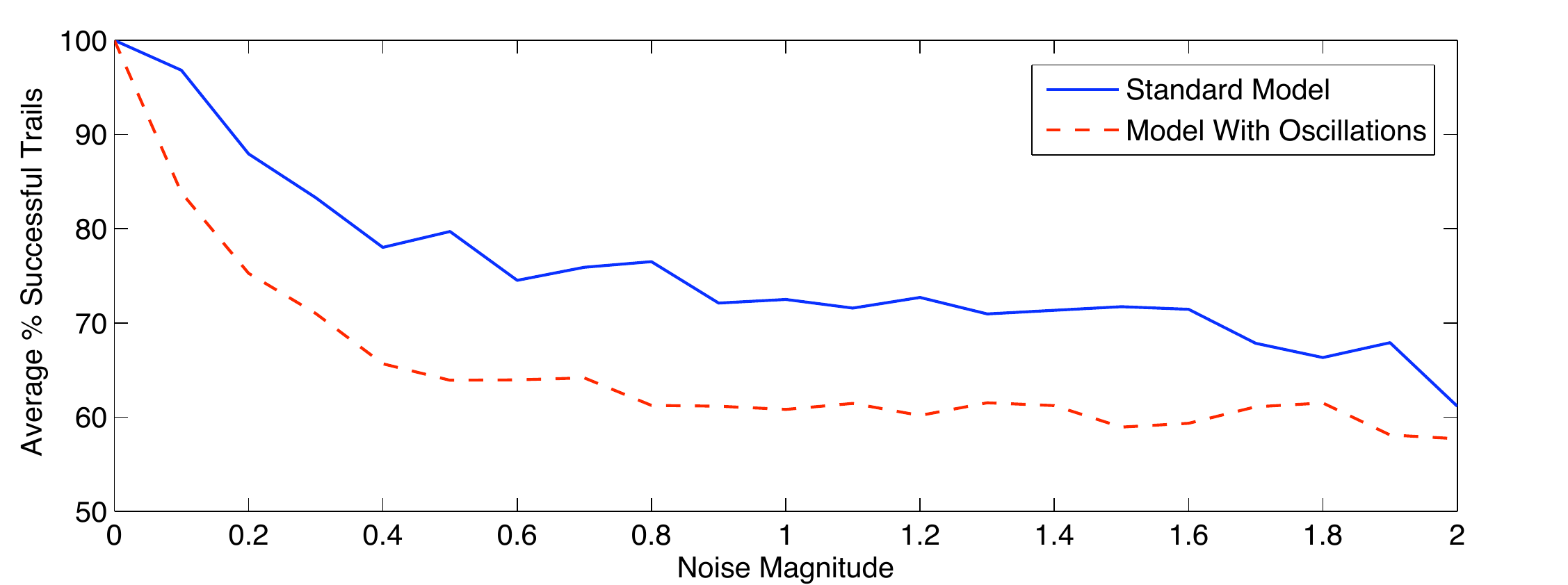}}\\
\subfloat[\textbf{Standard Deviation}]{
\includegraphics[width=5.6in]{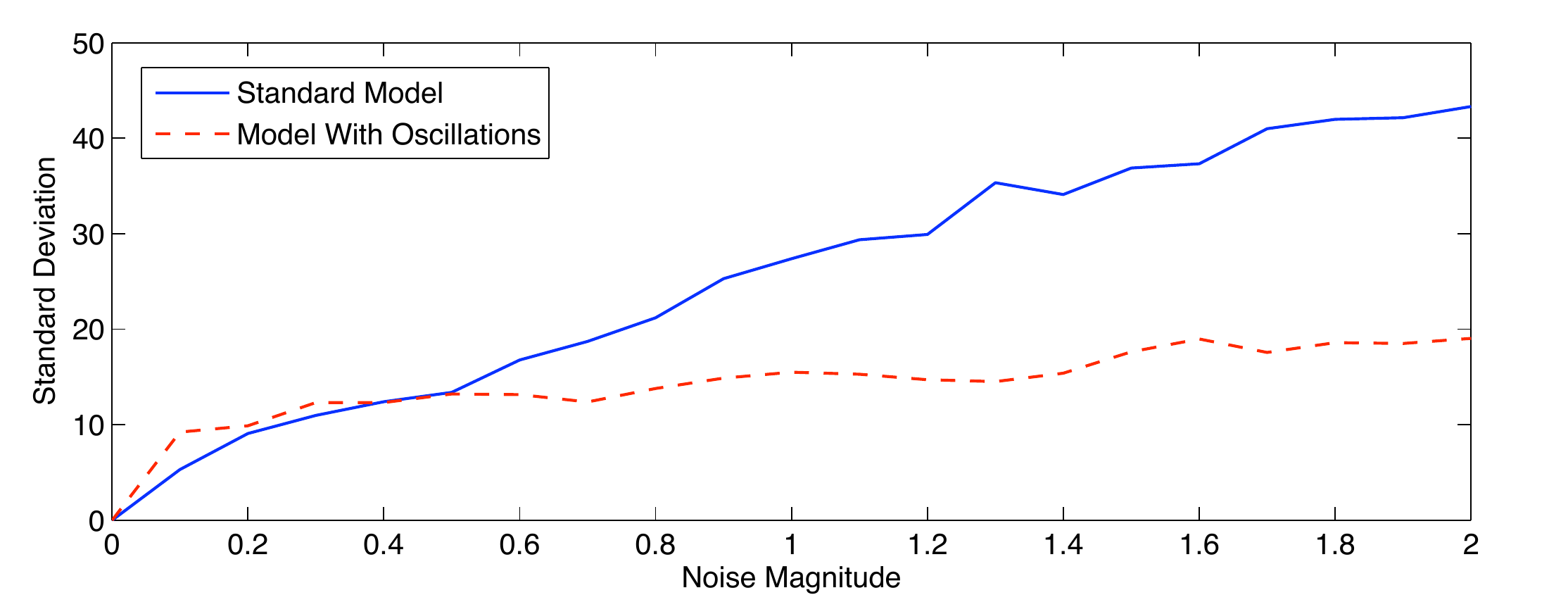}}
\caption{\textbf{Influence of Noise on Standard and Oscillating Model} - Comparison of average performance and standard deviation for the standard model (solid blue line) and oscillating model (dotted red line). For each noise magnitude, 200 simulation runs of 1000 seconds each were performed. \\ \vspace{-1.2cm}}\label{fig:ResStdModelError}
\end{figure}

In an attempt to understand why oscillations may be beneficial, tests were carried out to see the impact that noise had on each type of model. Noise magnitude was varied between $0$ and $2$ in $0.1$ intervals. For each magnitude 200 simulations of 1000 seconds long were run and average success rates computed for each simulation and across simulations sets. Also, to understand the spread of performance values, the standard deviation of the average success rate was calculated between simulation sets. Results can be found in Figure \ref{fig:ResStdModelError} and Table \ref{tab:ModelNoiseResults}.

The average success rate for both models falls quickly as noise is increased to 0.5, after which the rate becomes more stable. The oscillating model for the majority of noise magnitudes remains approximately $10$\% less than the standard model, however, that amount begins to become reduced after noise reaches 1.7.

These findings would appear to show that oscillation are in fact detrimental to overall performance. This is not, however, the full picture. If you compare the changes in standard deviation shown in Figure \ref{fig:ResStdModelError}(b), the standard model increases quickly in a linear manner while the oscillating model grows much slower after noise reaches a magnitude of 0.3. This leads to the standard model having considerably larger standard deviation after a noise magnitude of 0.5.

There are two reasons these results are interesting. Firstly, intuition may lead you to believe that each choice turn can be considered independent with a set probability based on the parameters that have been selected. This would allow the process to be modelled using a Binomial distribution where mean and standard deviation are linked based on the choice probability. For both cases though the standard deviations based on probabilities calculated from the means do not match those we see in the simulations. This implies that each choice turn is not an independent event, with the previous actions of the rat having an influence on the success probability. In some ways this effect could be viewed as a very simple form of short term learning where recent pervious events `prime' the system state such that an altered choice probability is formed. It does not, however, fit with standard learning models because no persistent long term changes are made to the system. Other than state changes, which are naturally transient due to the movement of the rat, the behaviour is caused due to an artefact of the stimuli structure (the maze shape) and the rules that the hard wired connections in the model relate to. 

This finding may also have wider implications when testing such models on a fixed structure problem. As we have used the same starting conditions for integrator states, it is possible that this could lead to an artificial performance that is dictated by the maze layout. To ensure that this is not the case in future simulations it may be preferable to randomise integrator states or run the models on mazes whose structure is not fixed, containing a random aspect. Unfortunately time did not permit the investigation of such scenarios, however, this would be an interesting extension. It also would require the modification of experimental configurations to allow comparison to real world results.

The second interesting result is that oscillations appear to reduce the growth of the standard deviation. This could be a factor in favour of using oscillations. If it is beneficial for consistent results over a wide range of noise magnitudes, then oscillations would help, having a similar distribution for noise magnitudes from 0.4 to 1.8. The standard model on the other hand would give a distribution that becomes more spread as noise is increased.

\newpage
\begin{table}[h]
\centering
\begin{tabular}{ c c c c c }
	\toprule
	\textbf{Noise Magnitude} & \multicolumn{2}{c}{\textbf{Average Success \%}} & \multicolumn{2}{c}{\textbf{Standard Deviation}}\\
	~ & Standard & Oscillating & Standard & Oscillating \\ 
	\midrule 
	0.0 & 100  & 100  & 0.0   & 0.0 \\
	0.1 & 96.8 & 83.8 & 5.3   & 9.2 \\
	0.2 & 87.9 & 75.2 & 9.1   & 9.9 \\
	0.3 & 83.2 & 71.0 & 11.0  & 12.3 \\
	0.4 & 78.0 & 65.7 & 12.4  & 12.3 \\
	0.5 & 79.7 & 63.9 & 13.4  & 13.2 \\
	0.6 & 74.5 & 64.0 & 16.8  & 13.1 \\
	0.7 & 75.9 & 64.2 & 18.7  & 12.4 \\
	0.8 & 76.5 & 61.3 & 21.2  & 13.8 \\
	0.9 & 72.1 & 61.2 & 25.3  & 14.9 \\
	1.0 & 72.5 & 60.8 & 27.4  & 15.5 \\
	\bottomrule
\end{tabular}
\caption{\textbf{Influence of Noise on Standard Model Results} - Detailed values for the data shown in Figure \ref{fig:ResStdModelError} for noise magnitudes up to 1. \\ \vspace{-1.2cm}}\label{tab:ModelNoiseResults}
\end{table}

\cleardoublepage
\chapter{Discussion}

\section{Conclusions}
A set of MATLAB scripts have been developed to help with the analysis of cross-frequency coupling in LFP recordings, focusing on phase to amplitude relations. These provide functionality to decompose the phase and amplitude of an input signal, calculate the ESC measure between two inputs and visualise the output as a heat plot to highlight possible areas of high correlation. All methods have been developed to allow for whole signals to be analysed as a single entity or through the use of a user specified moving window.

These methods were applied to real LFP recordings from the hippocampus and pre-frontal cortex, to compare any phase to amplitude coupling that occurs when approaching a choice and forced turn point in a T-maze task. Results from this analysis showed an increased ESC measure between the 1-10Hz hippocampus theta band and the 30-40Hz pre-frontal cortex gamma band for prolonged periods before the choice turn. The same levels of correlation were not present when approaching the forced turn and may point towards some form of co-ordinated action between these regions.

A standard connectionist model was produced that displayed performance results similar to real-world experiments. This was extended using shared inhibitory units which included a connection delay, to allow for oscillations to arise naturally. Experimentation with the simulations showed two types of dynamic being exhibited. A static stable state where the shared inhibitory units stopped any decision being made, and an oscillating state where the inhibitory units excitation was maintained between two levels while continually fluctuating. Although no oscillations were seen in the actual output or working memory integrators, only a limited set of parameter values were tested. Further tests would need to be performed to see if oscillations could be generated in these parts of the system.

A comparison was made between the standard and oscillating model in relation to resilience to noise. This illustrated that the standard model gave the best average performance, however, the standard deviation of the results distribution grew with noise. In contrast, the oscillating model maintained a similar standard deviation for a large range of noise magnitudes which may be a benefit if less accurate yet consistent behaviour is required.

Overall, the work has provided some backing to the use of cross-frequency coupling in the brain, presents a preliminary approach on how existing models could be altered to consider oscillatory components and highlights a possible reason they may be beneficial.

\section{Future Extensions}
Unfortunately due to limited time, the scope for this project had to be restricted. The following section details some possible extensions that could be carried out if the project was taken further.

\subsection{Phase Properties of Modelled Oscillations}
Now that it has been shown possible to incorporate oscillations into a standard connectionist approach, the next step is investigating possible ways in which these could be utilised. One such use would be for phase information to help bias decisions. For example, consider the scenario where we have each of our output integrators oscillating at a constant frequency, however, at shifted phases. On being activated by stimuli the amplitude of oscillations would increase but the frequency would remain constant. By setting the threshold for these integrators above the maximum amplitude of the oscillations and also allowing for the working memory to oscillate at the same frequency, it is possible for the threshold to only be exceeded when the phase of the related working memory and output integrators match. By controlling the phase of the working memory based on previous events, it is then possible to bias the decision made through constructive interference between the two signals. This type of method would be beneficial as it would allow for a much greater level of noise for the same task accuracy. As thresholds can be increased to much greater levels, while still being able to be reached, the chance of random activation due to noise is reduced. 

Although this type of idea seems promising there are many aspects that would need further consideration for a model like this to be created. These includes, a way of producing reliable oscillations at given frequencies and ways in which the phase of these can be shifted based on stimuli inputs. Even so it provides an interesting direction in which to take this work. 

\subsection{Consideration of Brain Connectivity}
The models developed so far were created in isolation to any real-world constraints. With connectivity and delays being vital components, it would useful to investigate the actual connectivity within the brain to see if realistic delays and connectivities give differing results. This may also help to answer questions as to why certain frequencies arise in actual brain recordings.

\cleardoublepage
\addcontentsline{toc}{chapter}{Acknowledgements}
\chapter*{Acknowledgements}

This work would have not been possible if not for the relentless patience of both Matt and Rafal, while I became familiar with neuroscience and modelling techniques outside of my previous experiences. I would also like to personally thank them both for providing so much of their time to sit down with me and ensure that the fundamental ideas from experimental and theoretical perspectives were fully understood.

\cleardoublepage
\renewcommand{\bibname}{References}
\addcontentsline{toc}{chapter}{References}
\bibliographystyle{plain}
\bibliography{Cross_Freq_Coupling}

\begin{thebibliography}{1}

\bibitem{Bruns:2004p9011}
A~Bruns and R~Eckhorn.
\newblock Task-related coupling from high-to low-frequency signals among visual
  cortical areas in human subdural recordings.
\newblock {\em International Journal of Psychophysiology}, 2004.

\bibitem{Buzsaki:2006}
G~Buzs\'{a}ki.
\newblock Rhythms of the brain.
\newblock Oct 2006.

\bibitem{LHahn:1996p9121}
Stefan~L. Hahn.
\newblock Hilbert transforms in signal processing.
\newblock Jan 1996.

\bibitem{Jones:2005p7436}
M~Jones and M~Wilson.
\newblock Theta rhythms coordinate hippocampal-prefrontal interactions in a
  spatial memory task.
\newblock {\em PLoS Biol}, Jan 2005.

\bibitem{Papoulis:1991}
A~Papoulis.
\newblock Probability, random variables, and stochastic processes.
\newblock 1991.

\bibitem{Penny:2008p8269}
W~Penny, E~Duzel, K~Miller, and J~Ojemann.
\newblock Testing for nested oscillation.
\newblock {\em Journal of Neuroscience Methods}, Jan 2008.

\bibitem{Singer:1999p7559}
Wolf Singer.
\newblock Neuronal synchrony: A versatile code for the definition of relations?
\newblock {\em Neuron}, Jan 1999.

\bibitem{Tort:2008}
A~B~L Tort, M~A Kramer, C~Thorn, D~J Gibson, Y~Kubota, A~M Graybiel, and N~J
  Kopell.
\newblock Dynamic cross-frequency coupling of local field potential
  oscillations in rat striatum and hippocampus during performance of a t-maze
  task.
\newblock {\em PNAS}, Dec 2008.

\bibitem{Varela:2001p7792}
F~Varela, J~Lachaux, E~Rodriguez, and J~Martinerie.
\newblock The brainweb: phase synchronization and large-scale integration.
\newblock {\em Nature Reviews Neuroscience}, Jan 2001.

\end{thebibliography}

\end{document}